\documentclass[aps,prd,psfig,amsmath,amsfonts,groupedaddress,eqsecnum]{revtex4}

\newcommand{\fourvec}[1]{\mbox{\boldmath $\mathsf{#1}$}}
\newcommand{\beq}{\begin{equation}}
\newcommand{\eeq}{\end{equation}}
\newcommand{\bea}{\begin{eqnarray}}
\newcommand{\eea}{\end{eqnarray}}
\newcommand{\bes}{\begin{subequations}}
\newcommand{\ees}{\end{subequations}}
\newcommand{\bari}{{I}\!\!\!{\scriptscriptstyle{{}^{\sim}}}}

\begin{document}



\title{Post-Newtonian gravitational radiation
and equations of motion via direct
integration of the relaxed Einstein equations. \\
III. Radiation reaction for binary systems with spinning bodies}


\author{Clifford M. Will}
\email[]{cmw@wuphys.wustl.edu}
\homepage[]{wugrav.wustl.edu/people/CMW}
\affiliation{Groupe Gravitation Relativiste et Cosmologie (GR$\varepsilon$CO)\\
Institut d'Astrophysique, 98 bis Boulevard Arago, 75014 Paris, France}
\affiliation{McDonnell Center for the Space Sciences, Department of
Physics, \\Washington University, St. Louis, Missouri 63130
\footnote{Permanent address}}


\date{\today}

\begin{abstract}
Using post-Newtonian equations of motion for fluid bodies that include
radiation-reaction terms at 2.5 and 3.5 post-Newtonian (PN) order
($O[(v/c)^5]$ and $O[(v/c)^7]$ beyond Newtonian order), we derive the
equations of motion for binary systems with spinning bodies.  In
particular we determine the effects of radiation-reaction coupled to
spin-orbit effects on the two-body equations of motion, and on the
evolution of the spins.  For a suitable 
definition of spin, we reproduce the standard equations of motion and
spin-precession at the first post-Newtonian order.  At 3.5PN order, we
determine the
spin-orbit induced reaction effects on the orbital motion, but 
we find that radiation damping has {\em no} effect on
either the magnitude or the direction of the spins.  Using the
equations of motion, we find that the loss of total energy and total angular
momentum induced by spin-orbit effects
precisely balances the radiative flux of 
those quantities calculated by
Kidder {\em et al.}  The equations of motion may be useful for
evolving inspiraling orbits of compact spinning binaries.
\end{abstract}


\maketitle

\section{Introduction and Summary }
\label{intro}

This is the third in a series of papers on motion and gravitational
radiation in the post-Newtonian approximation to general relativity (GR).
In  Paper I \cite{dire1}, we developed a method of direct integration of the
relaxed Einstein equations, whereby we wrote the Einstein equations
as an inhomogeneous flat-spacetime  wave equation together with a harmonic
gauge condition and solved them formally in terms of retarded integrals
over the past null cone of a given field point.  We obtained formal solutions
both for the far-zone gravitational waves and for the near-zone fields 
needed to obtain equations of motion.
We then developed the near-zone fields in a post-Newtonian expansion, in
powers of 
$\epsilon \sim (v/c)^2 \sim Gm/rc^2$, where $v$, $m$ and $r$ represent
typical velocities, masses and separations in the system, and $G$ and $c$
are the gravitational constant and speed of light (both set equal to unity
henceforth).  Each power of
$\epsilon$ represents one
``post-Newtonian'' (PN) order in the series ($\epsilon^{1/2}$ represents
one-half, or 0.5PN orders).  The near-zone metric was evaluated through
3.5PN order in terms of instantaneous, Poisson-like integrals over
distributions of perfect fluid.

In Paper II \cite{dire2}
we specialized the equations of motion to binary systems of
non-rotating, suitably spherical balls of pressureless fluid, whose size is
small compared to their separation, and derived the two-body equations of
motion through 2.5PN order and including the 3.5PN terms (calculation of
the 3PN terms is ongoing).  Through 2PN order, the equations of motion are
conservative, and our results were in complete agreement with calculations
of others\cite{DD81,damour300,kopeikin85,GK86,bfp98,futamase01}.  The 2.5PN and 3.5PN terms represent the 
effects of gravitational
radiation damping and their post-Newtonian corrections.  
We showed that they lead to losses of orbital energy and
angular momentum that correctly match the losses calculated from the
gravitational-wave flux to infinity~\cite{iyerwill,iyerwill2}.  They
are also consistent with a recent calculation of 3.5PN terms in the
equations of motion using the
post-Minkowskian approach \cite{luc3.5}.

In addition to the formal question of deriving suitably well-defined
equations of motion in general relativity, a problem that dates back to the
earliest days of the subject, this work is motivated by practical
considerations.  The operation of a network
of ground-based, kilometer-scale gravitational-wave
interferometers, and the research and development toward 
a space-based interferometer have
made it critical to obtain highly accurate theoretical models for the
evolution of and gravitational wave emission from two-body systems.  One of
the leading candidate sources for detection by interferometers
is the inspiral of a binary
system containing black holes or neutron stars, and the preferred method of
detection, optimal matched filtering, requires theoretical ``template''
waveforms that are accurate to fractions of a wave cycle over the
potentially thousands of cycles in a ``chirp'' signal whose frequency and
amplitude increase with time as the binary inspirals.  This is why
calculations to high PN order are needed.

However, the two-body equations of motion of Paper II (and indeed of all
high-PN order calculations done to date) have a shortcoming -- they
assume non-spinning bodies.  But spin may play a critical role in binary
inspiral, particularly involving black holes.  Spin-orbit and spin-spin
coupling leads to precessions of the spins of the bodies and of the
orbital plane, the latter effect 
resulting in modulations of the amplitude of the
gravitational waveform received at a detector.  Furthermore, spin effects
contribute directly to the gravitational waveform, and
to the overall emission of energy and angular momentum from the 
system~\cite{kww,kidder}. 

From the point of view of gravitational-wave data analysis, spin complicates
matters by increasing the number of parameters that must be estimated in a
binary-inspiral matched filter; this can significantly decrease the accuracy
with which any parameters can be measured, and can increase the
computational burden \cite{poissonwill,CC,SH,vecchio,bbw}.

Thus, it is desirable to have as complete a theoretical picture as is
reasonable of the effects of spin in relativistic binaries.  Formally, spin
effects first enter at the 1PN level.  By analogy with quantum
mechanics, the spin orbit contribution to the energy is of order ${\bf L}
\cdot {\bf S} /r^3$, where ${\bf L}$ is the orbital angular momentum.  This
can be re-expressed roughly in the form $\delta E \sim (mrv)(mRV)/r^3 \sim
(m^2/r)(R/r)vV \sim
\epsilon E_N $, where $R$ and $V$ are the
radius and rotational velocity
of the spinning body, and $E_N \sim m^2/r$ is the Newtonian orbital energy
\cite{PNnote}.  
Indeed, the 1PN effects of spin have been derived by
numerous authors from a variety of points of view, ranging from formal
developments of the GR equations of motion in multipole expansions
\cite{papapetrou1,papapetrou2}, to post-Newtonian calculations
\cite{obrien}, to
treatments of linearized GR as a spin-two quantum theory, with the concomitant
spinning-body interaction potentials \cite{barkerocon1,barkerocon2}.  
For a review of these various approaches, see \cite{barkeroconrev}.

The resulting two-body equations of motion can be written in the form 
\begin{equation}
{\bf a} =  -\frac{m}{r^2} {\bf n}+ {\bf a}_{\rm PN} + {\bf a}_{\rm SO} 
+ \dots \,,
\label{eomsummary}
\end{equation}
where ${\bf a} = {\bf a}_1 - {\bf a}_2$ is the relative acceleration,
and where the 1PN point mass and spin-orbit contributions are given by
\begin{subequations}
\begin{eqnarray}
{\bf a}_{\rm PN}  &=& -\frac{m}{r^2} \biggl \{ {\bf n} \left 
[ (1 + 3\eta) v^2 - 2 (2+\eta) \frac{m}{r} - \frac{3}{2} \eta {\dot
r}^2 \right ]
-2 (2-\eta) \dot r {\bf v} \biggr \} \,,
\label{eomsummaryPN}
\\
{\bf a}_{\rm SO}  &=& \frac{1}{r^3} \biggl \{
\frac{3}{2} \frac{\bf n}{r} {\bf {\tilde L}}_{\rm N} \cdot 
\left ( 4{\bf {\cal S}} + 3\fourvec{\xi} \right )
- {\bf v} \times 
\left ( 4{\bf {\cal S}} + 3\fourvec{\xi} \right )
+\frac{3}{2} \dot r {\bf n} \times 
\left ( 4{\bf {\cal S}} + 3\fourvec{\xi} \right )
\biggr \} \,,
\label{eomsummarySO}
\end{eqnarray}
\label{eomsummary2}
\end{subequations}
where ${\bf x} \equiv {\bf x}_1 - {\bf x}_2$, $r \equiv |{\bf x}|$,
${\bf n}
\equiv {\bf x}/r$, $m \equiv m_1 + m_2$, $\eta \equiv m_1m_2/m^2$,
${\bf v} \equiv {\bf v}_1 - {\bf v}_2$, 
$\dot r = dr/dt = {\bf n} \cdot  {\bf v}$, ${\bf {\cal S}} = {\bf {\cal S}}_1 + {\bf {\cal S}}_2$,
$\fourvec{\xi} = (m_2/m_1){\bf {\cal S}}_1 + (m_1/m_2){\bf {\cal S}}_2$, and 
${\bf {\tilde L}}_{\rm N} = {\bf x} \times {\bf v}$.
The 1PN equations of spin precession are given by
\begin{subequations}
\bea
{\dot {\bf {\cal S}}}_1 &=& \frac{\eta m}{r^3} {\bf {\tilde L}}_{\rm N} 
\times {\bf {\cal S}}_1 \left ( 2 + \frac{3}{2}\frac{m_2}{m_1} \right ) \,, 
\\
{\dot {\bf {\cal S}}}_2 &=& ( 1 \rightleftharpoons 2 )\,.
\eea
\label{spinsummary}
\end{subequations}
These equations assume a specific definition of the center of mass and spin
of each
spinning body, given by
${\bf x}_1 \equiv m_1^{-1} \int_1 \rho^* {\bf x} d^3x$, 
and
${\bf {\cal S}}_1 \equiv {\bf S}_1 (1+\frac{1}{2}v_1^2 +3m_2/r) -
\frac{1}{2}{\bf v}_1 \times ({\bf v}_1 \times {\bf S}_1)$,
with analogous expressions for body 2,
where $\rho^*$ is the so-called ``conserved'' or baryonic density [Eq.
(\ref{rhostar})], $m_1$ is
the baryonic mass, and ${\bf S}_1 \equiv \int_1 \rho^* {\bf {\bar x}} \times
{\bf {\bar v}} d^3x$, is the baryonic spin,
with ${\bf {\bar x}} = {\bf x}-{\bf x}_1$.
The
spin orbit terms in Eq. (\ref{eomsummarySO}) and the spin precession
equations (\ref{spinsummary}) derived by our method
are in complete agreement with earlier derivations \cite{barkeroconrev}.

Turning now to radiation-reaction effects, it is clear that no spin
contributions can occur at the 2.5PN order corresponding to the
leading-order damping effects.  The simplest way to see this is to recall
that, in a certain gauge, the 2.5PN radiation reaction acceleration may be
expressed in the form 
\beq
a_{\rm 2.5PN}^i = -\frac{2}{5} x^j \frac{d^5}{dt^5} \bari^{ij}  \,,
\eeq
where $\bari^{ij}$ is the trace-free moment of inertia of the system, and
where we sum over repeated spatial indices.  But
since, to lowest order, $\bari^{ij}$ contains no velocities, it cannot contain
spin terms at lowest 
order, even if the bodies are rotating; in other words, spin
terms will
be of higher, 3.5PN order.  Similarly, in calculating
${\dot {\bf S}}_1$ from ${\bf x} \times {\bf a}_{\rm 2.5PN}$ it is also clear
that no spin terms can arise at lowest order.  
Hence, the leading contributions of spin in
radiation reaction must be at 3.5PN order. 

We therefore make use of the 2.5PN {\em and} 3.5PN equations of motion for fluid
systems derived in Paper I and calculate the equations of motion and spin
precession for two spinning bodies.  
These are our central results:
the 2.5PN contributions to radiation reaction are given by the
standard ``point-mass'' result
\beq
{\bf a}_{\rm 2.5PN} = \frac{8}{5} \eta \frac{m^2}{r^3} 
\left [ {\dot r}{\bf n} \left ( 3v^2 + \frac{17}{3}\frac{m}{r} \right )
-{\bf v} \left ( v^2 +3\frac{m}{r} \right ) \right ]
\,,
\label{arel2.5PN}
\eeq
and
the spin-orbit
contributions to radiation reaction are given by
\bea
{\bf a}_{\rm 3.5PN-SO} &=& 
-\frac{\eta m}{5r^4} 
\biggl \{
\frac{{\dot r}{\bf n}}{r}
\left [ \left ( 120v^2+280{\dot r}^2+453\frac{m}{r} \right )
{\bf {\tilde L}}_{\rm N} \cdot {\bf {\cal S}}
\right .
\nonumber \\
&&
\left .
+ \left ( 285v^2-245{\dot r}^2+211\frac{m}{r} \right )
{\bf {\tilde L}}_{\rm N} \cdot \fourvec{\xi}
\right ]
\nonumber \\
&&
+
\frac{\bf v}{r}
\left [ \left ( 87v^2-675{\dot r}^2-\frac{901}{3}\frac{m}{r}
\right )
{\bf {\tilde L}}_{\rm N} \cdot {\bf {\cal S}}
+ 4\left ( 6v^2-75{\dot r}^2 - 41\frac{m}{r} \right )
{\bf {\tilde L}}_{\rm N} \cdot \fourvec{\xi}
\right ]
\nonumber \\
&&
- \frac{2}{3}{\dot r}{\bf v} \times {\bf {\cal S}}
\left ( 48v^2 + 15{\dot r}^2+364\frac{m}{r} \right )
- \frac{1}{3}{\dot r}{\bf v} \times \fourvec{\xi}
\left ( 375v^2 -195{\dot r}^2+640\frac{m}{r} \right )
\nonumber \\
&&
+\frac{1}{2}{\bf n}\times {\bf {\cal S}}
\left ( 31v^4-260v^2{\dot r}^2+245{\dot r}^4
-\frac{689}{3}v^2\frac{m}{r} + 537{\dot r}^2\frac{m}{r}
+\frac{4}{3}\frac{m^2}{r^2} \right )
\nonumber \\
&&
-\frac{1}{2}{\bf n}\times \fourvec{\xi}
\left ( 29v^4-40v^2{\dot r}^2-245{\dot r}^4
+211v^2\frac{m}{r} - 1019{\dot r}^2\frac{m}{r}
-80\frac{m^2}{r^2} \right )
\biggr \} \,.
\label{aSORR}
\eea
The point-mass, 3.5PN contributions are as given in Paper II, Eqs. (1.2) and
(1.3d).
We also find the radiation-reaction contribution to spin precession
\beq
{\dot {\bf {\cal S}}}^{\rm 3.5PN-SO}_1 = 0 \,, \quad
{\dot {\bf {\cal S}}}^{\rm 3.5PN-SO}_2 = 0 \,.
\label{spinrr}
\eeq
Although the latter result arises effectively
from a cancellation of many terms, it
should really come as no surprise.  Our bodies are assumed to be axially
symmetric, and as such, should not couple to 
gravitational radiation on the basis
of their rotation alone.  Their
{\em precessions} could lead to gravitational radiation and to
back-reaction if, for example, they
were rotationally flattened, but those precessions and the flattening are
themselves
1PN-order effects, and so should result in 
radiation reaction effects of at least 4.5PN order, beyond the level
at which we are working.  

We have verified that
these equations lead to a loss of total energy and
total
angular momentum that matches precisely the energy and angular momentum flux
at infinity from binary systems of spinning bodies, as calculated by Kidder
{\em et al.}~\cite{kww,kidder}.  

In the conventional terminology of spinning bodies in general relativity, 
the center of mass of each body is specified by a so-called
``spin supplementary
condition'' (Appendix \ref{app:ssc}), given by a parameter $k_{\rm SSC}$ whose
value is typically zero, one or 1/2.  Our center-of-mass definition 
corresponds to the value $k_{\rm SSC}=1/2$.  Appendix \ref{app:ssc=1}
displays the final equations of motion corresponding to the ``covariant'' SSC
given by $k_{\rm SSC}=1$. 

The remainder of this paper provides details.  
In Sec. \ref{sec2.5pn} we derive the
equations of motion and spin precession to 1PN and 2.5 PN order, and demonstrate
explicitly the absence of observable radiation-reaction
spin effects at 2.5PN  order.
In Sec. \ref{sec3.5pn} we complete the derivation to 3.5PN order, and verify
the agreement with energy and angular momentum flux.  A variety of
technical matters and detailed equations are relegated to 
Appendices.

\section{Post-Newtonian and 2.5PN equations of motion}
\label{sec2.5pn}

\subsection{Foundations}
\label{foundations}

We wish to analyse a binary system consisting of balls of perfect fluid that
are sufficiently small compared to their separation that tidal interactions
(and their relativistic generalizations) can be ignored, but that are
sufficiently extended that they can support a finite rotational angular
momentum, or spin.  At Newtonian order, the result is essentially trivial:
the equation of motion for body 1 is $d^2 {\bf x}_1/dt^2 = -m_2 {\bf x}/r^3 +
O(mR^2/r^4)$, where $R$ is the characteristic size of the bodies.  Spin
plays no role whatsoever, because the Newtonian interaction does not depend
on velocity.   But at post-Newtonian order, there are velocity-dependent
accelerations of the schematic form $mv^2/r^2$, and thus, taking into
account the finite size of the body and expanding about its center of mass, 
one could anticipate acceleration terms
of the form $(mVR)v/r^3 \sim Sv/r^3$.  
However, the combination of finite size and spin introduces an
ambiguity in the definition of the center of mass of each body.  At
Newtonian order, the center of mass is defined naturally by ${\bf x}_A =
m_A^{-1}\int_A \rho {\bf x} d^3x$, where $\rho$ is the density and where
the integral is over the body in
question.  But at post-Newtonian order, depending on the choice of
``density'' used, there could be correction terms in the center of mass
of the form $m^{-1}\int
\rho v^2 {\bf x} d^3x \sim vS/m$.  Because ``center of mass'' is an
arbitarily chosen point, these differences have no physical content in the
end, as long as results are expressed in terms of measurable quantities
(such as the total energy of the system, or the radar range to the surface
of the body), but they do result in equations of motion with different
explicit forms.  This has given rise to the concept of ``spin supplementary
condition'' (SSC), a statement about which center of mass definition
is being used; this concept is discussed in Appendix \ref{app:ssc}.

Because we intend to work to post-Newtonian orders that include 1PN, 2.5PN
and 3.5PN terms, we need to define centers of mass and spins in a way that
is physically reasonable, calculationally simple, and easily transformable
to definitions based on other criteria.  In post-Newtonian theory, the
simplest ``density'' available is the so-called ``conserved'' density, given
by
\begin{equation}
\rho^* \equiv \rho \sqrt{-g} u^0 \,,
\label{rhostar}
\end{equation}
where $\rho$ is the mass energy density as measured by an observer in a
local inertial frame momentarily at rest with respect to the fluid, $g$ is
the determinant of the metric, and $u^0$ is the time component of the fluid
four-velocity.  Making the reasonable assumption that $\rho$ is directly
proportional to the baryon number density, then, by virtue of conservation
of baryons, we have that $(\rho u^\alpha)_{; \alpha} =0$, where semicolon
denotes covariant derivative, and the index $\alpha$ is summed over all
spacetime values.  This is
equivalent to the {\it exact} continuity equation
\begin{equation}
\partial \rho^* /\partial t + \nabla \cdot  (\rho^* {\bf v}) = 0 \,,
\label{continuity}
\end{equation}
where $v^i = u^i/u^0$ is the ordinary (coordinate) velocity of the fluid.
The use of $\rho^*$ in defining such quantities as center of mass is
convenient because of the fact, based on the equation of continuity, that 
\begin{equation}
\frac{\partial}{\partial t} \int \rho^* (t,{\bf x}^\prime)
f({\bf x},{\bf x}^\prime) d^3x^\prime = \int \rho^* 
(t,{\bf x}^\prime){\bf v}^\prime
\cdot \nabla^\prime f({\bf x},{\bf x}^\prime) d^3x^\prime \,,
\end{equation}
for any suitably regular function $f({\bf x},{\bf x}^\prime)$ and for
integration over a complete body.

Accordingly, we will define the baryonic
mass, center of mass and baryonic spin of each body in our
system to be
\bes
\begin{eqnarray}
m_A &\equiv&\int_A \rho^*d^3x \,,
\label{baryonmass}
\\
{\bf x}_A &\equiv& m_A^{-1} \int_A \rho^* {\bf x} d^3x \,,
\label{baryoncenter}
\\
{\bf S}_A &\equiv& \epsilon^{ijk} \int_A \rho^* {\bar x}^j {\bar v}^k d^3x \,, 
\label{baryonspin}
\end{eqnarray}
\label{baryonxs}
\ees
where ${\bar x}^j = x^j - x_A^j$ and ${\bar v}^j = v^j - v_A^j$.   For
future use we will also define a tensorial spin quantity
\begin{eqnarray}
S_A^{ij} &\equiv 2& \int_A \rho^* {\bar x}^{[i}{\bar v}^{j]} d^3x \,,
\nonumber \\
S_A^{ij} &=& \epsilon^{ijk} S_A^k \,,\qquad S_A^i = \frac{1}{2} \epsilon^{ijk}
S_A^{jk} \,,
\label{spintensor}
\end{eqnarray}
where $[\,]$ around indices denotes antisymmetrization. 
We will demonstrate in Appendix \ref{app:ssc} 
that this definition of center of mass corresponds
to the SSC value $k_{\rm SSC}=1/2$.
With these definitions,  the baryonic mass $m_A$ is constant, and 
the velocity, acceleration and rate of change of
spin of body $A$ are given by
\bes
\begin{eqnarray}
{\bf v}_A &=& m_A^{-1} \int_A \rho^* {\bf v} d^3x \,,
\label{bodyv}
\\
{\bf a}_A &=& m_A^{-1} \int_A \rho^* {\bf a} d^3x \,,
\label{bodya}
\\
d S_A^i/dt &=& \epsilon^{ijk} \int_A \rho^* {\bar x}^j {a}^k
d^3x \,.
\label{bodysdot}
\end{eqnarray}
\label{bodyvas}
\ees
Note that, because $\int_A \rho^* {\bf {\bar x}} d^3x =0$, we do not need
the ``barred'' version of the acceleration in Eq. (\ref{bodysdot}).

\subsection{Baryonic equations of motion and spin precession}
\label{sec2.5pnA}

We begin by working to 1PN and 2.5PN order, reproducing a number of
well-known 1PN formulae for spinning bodies, and establishing some results that
will be useful when we go on to 3.5PN order.  Since we are only interested
in radiation-reaction aspects of spin, we can ignore the 2PN terms in the
equations of motion; these produce only PN corrections to the spin equations
of motion.  Incorporating the 2PN terms would require a more accurate
definition of spin, including PN corrections.  Such a framework has not been
developed to date and is beyond the scope of this paper.  Initially, we derive
everything in terms of our baryonic definitions of center of mass and spin,
then later, we transform to more relevant definitions.

We use the continuum equations of
motion derived in Paper II, Eqs. (2.23), (2.24a) and (2.24c), 
with all quantities expressed in terms of the
conserved density $\rho^*$.  They are given by   
\beq
d^2 x^i/dt^2 = U^{,i} + a_{\rm PN}^i + a_{\rm 2.5PN}^i \,,
\label{fluid}
\eeq
where
\bes
\bea
a_{\rm PN}^i &=&  v^2 U^{,i} -4v^iv^j  U^{,j}- 4 U U^{,i} - 3 v^i {\dot U}
+ 4{\dot V}^i + 8v^j V^{[i,j]} 
 \nonumber \\
&& 
+ \frac{3}{2} \Phi_1^{,i} - \Phi_2^{,i}
+\frac{1}{2} {\ddot X}^{,i} \,,
\label{fluidPN}
\\
a_{\rm 2.5PN}^i &=&  
\frac{3}{5} x^j ( \stackrel{(5)}{{\cal I}^{ij}} - \frac{1}{3}
\delta ^{ij} \stackrel{(5)}{{\cal I}^{kk}} ) +
2 v^j \stackrel{(4)}{{\cal I}^{ij}}
+ 2 U^{,j} \stackrel{(3)}{{\cal I}^{ij}}
+ \frac{4}{3} U^{,i} \stackrel{(3)}{{\cal I}^{kk}}
- X^{,ijk} \stackrel{(3)}{{\cal I}^{jk}} 
\,,
\label{fluid2.5PN}
\eea
\label{fluidterms}
\ees
where commas denote partial derivatives, overdots denote partial time
derivatives, and $(n)$ above quantities denotes
the number of total time derivatives.  
The potentials used here and elsewhere in the paper are given by
the general definitions
\bea
\Sigma (f) &\equiv& \int \frac{\rho^*(t,{\bf x}^\prime) 
f(t,{\bf x}^\prime)}{|{\bf x} - {\bf x}^\prime |}
d^3x^\prime \,,
\nonumber \\
X(f) &\equiv& \int \rho^*(t,{\bf x}^\prime) f(t,{\bf x}^\prime)
|{\bf x} - {\bf x}^\prime | d^3x^\prime \,,
\nonumber \\
Y(f)  &\equiv& \int \rho^*(t,{\bf x}^\prime) f(t,{\bf x}^\prime)
|{\bf x} - {\bf x}^\prime |^3 d^3x^\prime \,,
\label{genpotentials}
\eea
with specific potentials given by 
\bea
U &=& \Sigma(1) \,, \quad
V^i = \Sigma(v^i) \,,
\nonumber \\
\Phi_1 &=& \Sigma(v^2) \,, \quad
\Phi^{jk}_1 = \Sigma(v^j v^k) \,, \quad
\Phi_2 = \Sigma(U) \,,
\nonumber \\
X &=& X(1) \,,  \quad X^i  = X(v^i) \,, \quad X_1 = X(v^2) \,, \quad
X_2 = X(U) \,,
\nonumber \\
Y &=& Y(1) \,.
\label{potentials}
\eea
The multipole moments of the system ${\cal I}^{ij}$, 
${\cal I}^{ijk}$, and ${\cal J}^{qj}$, and additional moments that
will be relevant at 3.5PN order are defined in Appendix \ref{app:moments}.
In Eq. (\ref{fluid2.5PN}), we have eliminated terms that are purely
functions of time whose main effect is to generate an overall center of mass
motion for the system that is irrelevant for our purposes.  This is
discussed in Appendix \ref{app:fluid}. 

We now multiply the equation of motion (\ref{fluid}) by $\rho^*$ and 
integrate over body 1, expressing the variables ${\bf x}$ and ${\bf v}$ as
${\bf x}={\bf x}_A + {\bf {\bar x}}$
and
${\bf v}={\bf v}_A + {\bf {\bar v}}$, 
where $A = 1,\,2$, depending on the body in which the point lies.   
To get the acceleration of body 1, we divide the result by
$m_1$.  We use
Eqs. (\ref{baryonxs}) and (\ref{bodyvas}) to simplify where possible.  
We expand the various
potentials in powers of ${\bar x}/r$, and keep terms proportional to ${\bar
v} \times {\bar x}$, as well as ``internal'' terms proportional to ${\bar
v}^2$, and $\rho^*/{\bar r}$.  The random part of the ``${\bar
v}^2$'' terms can be thought of as pressure contributions to the internal
structure, so we no longer are treating the bodies as purely pressureless.
We make extensive use of virial
relations
derived in Appendix \ref{app:virial} to simplify expressions dependent on the
internal structure of each body.  We also assume that the bodies are
approximately spherically symmetric in their own rest frames (we ignore
centrifugal and Lorentz-boost
flattening, which will be  higher-PN-order effects), so that
integrals of internal expressions with an odd number of spatial
indices (corresponding to odd-$l$ spherical harmonics) can be
set to zero.  We also ignore internal body 
terms that vary as ${\bar x}^2$, which
represent quadrupole and higher ``tidal'' effects, and which vanish as the
bodies' sizes tend to zero.

The Newtonian term gives $a_{\rm N}^i = -m_2 x_{12}^i/r^3$, where, in this
paragraph, we denote $x_{12}^i \equiv x_1^i - x_2^i$, 
$r \equiv |{\bf x}_{12}|$, and ${\bf n} = {\bf x}_{12}/r$.  
To illustrate the approach, we show
two examples of the terms that arise  at 1PN order, namely
\bea
\frac{1}{m_1} \int_1 \rho^* v^2 U^{,i} d^3x &=& -\frac{1}{m_1} 
 \int_1 \rho^* (v_1^2 + 2{\bf v}_1 \cdot {\bf {\bar v}} + {\bar v}^2)
  d^3x
  \nonumber \\
  && \times
  \biggl [ \int_1 \frac{\rho^{*\prime} (x-x^\prime)^i}{|{\bf x} - {\bf
  x}^\prime |^3}d^3x^\prime 
  + \frac{m_2 x_{12}^i}{r^3} + \frac{m_2{\bar x}^j (\delta^{ij} -
     3n^in^j)}{r^3} + \dots \biggr ]
\nonumber \\
&=& \frac{2v_1^j}{m_1} {\cal H}_1^{ji} - m_2 v_1^2 \frac{x_{12}^i}{r^3}
     - \frac{2{\cal T}_1}{m_1} \frac{m_2 x_{12}^i}{r^3}
     - \frac{S_1^{jk}v_1^k}{m_1} \frac{m_2 (\delta^{ij}-3n^in^j)}{r^3} \,,
\eea
and
\bea
\frac{1}{m_1} \int_1 \rho^* UU^{,i} d^3x &=& 
-\frac{1}{m_1}
\int_1 \rho^* d^3x \biggl [ 
\int_1 \frac{\rho^{*\prime\prime}}{|{\bf x} - {\bf x}^{\prime\prime} |}
d^3x^{\prime\prime}
+\frac{m_2}{r} - \frac{m_2}{r^3} {\bf {\bar x}} \cdot {\bf x}_{12} + \dots
\biggr ]
\nonumber \\
&&  \times \biggl [  \int_1 \frac{\rho^{*\prime} (x-x^\prime)^i}{|{\bf x} 
-{\bf x}^\prime |^3}d^3x^\prime
+ \frac{m_2 x_{12}^i}{r^3} - \frac{m_2{\bar x}^j (\delta^{ij} -
     3n^in^j)}{r^3} + \dots \biggr ]
\nonumber \\
&&
= - \frac{\Omega_1^{ij}}{m_1}\frac{m_2 x_{12}^j}{r^3}
  + 2\frac{\Omega_1}{m_1}\frac{m_2 x_{12}^i}{r^3}
  - \frac{m_2^2 x_{12}^i}{r^4}
\,,
\eea
where 
${\cal T}_1$,
$\Omega_1^{ij}$, 
$\Omega_1$ and
${\cal H}_1^{ij}$ are defined in Appendix \ref{app:virial}.
These entirely internal, 
structure-dependent 
terms can be eliminated via virial relations; however they can and
do
generate spin terms at 1PN and 3.5PN order.  Detailed expressions for
these virial relations may be found in Appendix \ref{app:virial}.

In the combination of 1PN terms $ 4{\dot V}^i 
+\frac{1}{2} {\ddot X}^{,i}$ in Eq. (\ref{fluidPN}), 
the time derivatives generate accelerations
inside the potentials.
To the order needed for our purposes, we must therefore
substitute the Newtonian and
2.5PN continuum 
terms for those accelerations and carry out the same procedures for the
integrals as described
above.  Those Newtonian and 2.5PN acceleration
terms inserted into 1PN expressions will 
then generate respective 1PN and 3.5PN point mass and spin-orbit 
contributions.

Turning to the 2.5PN terms, Eq. (\ref{fluid2.5PN}),  
the integrations lead to no explicit internal or
spin terms; however for our 3.5PN accurate work, 
the multipole moments and their time derivatives must
be evaluated to 1PN order, and in those 1PN corrections, spin terms will
appear (see Appendix \ref{app:moments}).  

The resulting equation of motion for body 1 is
\beq
a_1^i = -\frac{m_2 n^i}{r^2} + (a_1^i)_{\rm PN} + (a_1^i)_{\rm SO} + 
(a_1^i)_{\rm 2.5PN} 
\,,
\label{a1summary0}
\eeq
where
\bes
\bea
(a_1^i)_{\rm PN} &=& \frac{m_2 n^i}{r^2} \left [ 5\frac{m_1}{r} +
4\frac{m_2}{r} - v_1^2 + 4{\bf v}_1 \cdot {\bf v}_2 -2v_2^2 
+ \frac{3}{2} ({\bf v}_2 \cdot {\bf n} )^2 \right ]
\nonumber \\
&& - \frac{m_2}{r^2} (v_1 - v_2)^i (3{\bf v}_2 \cdot {\bf n}
 - 4{\bf v}_1 \cdot {\bf n} ) \,,
 \label{a1PN}
\\
(a_1^i)_{\rm SO} &=&
\frac{m_2}{r^3} (\delta^{ij} - 3n^in^j)
\left [ 2(v_2^k {\tilde S}_1^{jk} - v_1^k {\tilde S}_2^{jk} )
-\frac{3}{2} (v_1^k {\tilde S}_1^{jk} - v_2^k {\tilde S}_2^{jk} )
\right ]
\nonumber \\
&&
-\frac{m_2}{r^3} (\delta^{jk} - 3n^jn^k)
(v_1 - v_2)^j (\frac{3}{2} {\tilde S}_1^{ik} +2 {\tilde S}_2^{ik} )
\,,
\label{a1SO}
\\
(a_1^i)_{\rm 2.5PN} &=& 
\frac{3}{5} x_1^j ( \stackrel{(5)}{{\cal I}^{ij}} - \frac{1}{3}
\delta ^{ij} \stackrel{(5)}{{\cal I}^{kk}} ) +
2 v_1^j \stackrel{(4)}{{\cal I}^{ij}}
- \frac{1}{3} \frac{m_2}{r^2} n^i \stackrel{(3)}{{\cal I}^{kk}}
- 3 \frac{m_2}{r^2} n^in^jn^k \stackrel{(3)}{{\cal I}^{jk}} 
\,,
\label{a12.5PN}
\eea
\label{a1summary}
\ees
where ${\tilde S}_A^{ij} \equiv {S}_A^{ij}/m_A$, and where the ``mass'' $m_2$
in
the Newtonian term is now given, to PN order, by the baryonic mass plus 
$\frac{1}{2}\Omega_2$, which corresponds to the total mass-energy of the body.  

We calculate the precession of the spin in a similar manner.  Starting with
$dS_1^i/dt = \epsilon^{ijk} \int_1 \rho^*{\bar x}^j a^k d^3x$, 
we expand about the
baryonic centers of mass, keeping terms that depend on ${\bar x} \times
{\bar v}$, dropping internal terms with odd numbers of spatial indices, and
throwing away terms that vary as $R^2$ or higher.  Notice that, even in
Newtonian theory, such $(m_2/r^3)(m_1 R_1^2)$ terms occur, and represent
standard quadrupole coupling; in the Earth-Moon system, these lead to the
precession of the equinoxes.  However, as we wish to deal with compact
bodies, we shall ignore such effects.  At 1PN order, the only terms in Eq.
(\ref{fluidPN}) that
contribute are those that have explicit velocity dependence.  The result, at
1PN order is 
\beq
({\dot S}_1^i)_{\rm PN} = \frac{m_2}{r^2} \biggl [ 5S_1^i ({\bf v}_1 \cdot {\bf n})
-3S_1^i ({\bf v}_2 \cdot {\bf n}) - 2n^i ({\bf S}_1 \cdot {\bf v})
+(v_1 -2v_2)^i ({\bf S}_1 \cdot {\bf n}) \biggr ] \,.
\label{spinprecess}
\eeq
Strangely, however, there {\em is} a 2.5PN contribution to the spin equation
of motion, arising from the $v$-dependent term in Eq. (\ref{fluid2.5PN}),
and given by
\beq
({\dot S}_1^i)_{\rm 2.5PN} = 
S_1^i \stackrel{(4)}{{\cal I}^{jj}} - S_1^j \stackrel{(4)}{{\cal I}^{ij}} \,.
\label{spin2.5PN}
\eeq
However, we
notice that, since the spin is constant to lowest order, then, to 2.5PN
order, the right-hand-side of Eq. (\ref{spin2.5PN}) is a total time
derivative, and thus can be moved to the left-hand-side and 
absorbed into a redefinition of the spin.  
In any case, 
it is clear from the argument made in Sec. \ref{intro} that this is a
spurious effect, because there is a gauge in which the 2.5PN terms in the
equation of motion have no explicit velocity dependence.  This 2.5PN gauge,
sometimes called Burke-Thorne gauge \cite{MTW}, may be obtained from our
gauge by the coordinate transformation
\bea
x^i &\to& x^i - x^j \stackrel{(3)}{{\cal I}^{ij}} + \frac{1}{3}
x^i \stackrel{(3)}{{\cal I}^{jj}} \,,
\nonumber \\
t &\to& t + \frac{2}{3} \stackrel{(2)}{{\cal I}^{jj}} \,.
\eea
Applying this transformation to our definition of baryonic spin
Eq. (\ref{baryonspin}), and recalling that $\rho^* d^3x$ is an invariant
quantity, we find that 
\beq
(S_1^i)_{\rm Burke-Thorne} = S_1^i - S_1^i \stackrel{(3)}{{\cal I}^{jj}} +
S_1^j \stackrel{(3)}{{\cal I}^{ij}} \,.
\label{spinburkethorne}
\eeq
Hence, absorbing the 2.5PN terms into the definition of the spin is
equivalent to defining our spin in the Burke-Thorne gauge.
Although this redefinition eliminates 2.5PN terms from the spin precession,
it {\em will} have consequences at 3.5PN order.  

\subsection{Post-Newtonian conserved quantities}

It is useful to verify that, at 1PN order, the equations of motion and spin
precession we have found admit suitable conserved quantities for energy, 
angular momentum and momentum or center-of-mass motion.  
Taking the equations of motion (\ref{a1summary0}),
(\ref{a1PN}) and (\ref{a1SO}), contracting with $m_1 {\bf v}_1$, summing
over both bodies, and using the equations of motion
to extract time derivatives (see Appendix \ref{app:timederiv}), 
one obtains a conserved total energy.
Doing the same procedure with a cross product with $m_1 x_1$, and combining
with the spin precession equations, one obtains a conserved total angular
momentum.  Finally, multiplying by $m_1$, summing over both bodies and
extracting time derivatives, one obtains an expression for the system center
of mass.  These 1PN-conserved quantities are given by
\bes
\bea
E &=&  
\frac{1}{2} m_1 v_1^2 - \frac{1}{2} \frac{m_1m_2}{r} 
 + \frac{3}{8} m_1 v_1^4 +  \frac{3}{2} m_1 v_1^2 \frac{m_2}{r}
+ \frac{1}{2} \frac{m_1m_2^2}{r^2}
\nonumber \\
&&
- \frac{1}{4} \frac{m_1m_2}{r} (7{\bf v}_1 \cdot {\bf v}_2
+ {\bf v}_1 \cdot {\bf n} \, {\bf v}_2 \cdot {\bf n} ) 
 + (1 \rightleftharpoons 2 ) \,,
\label{EPN}\\
{\bf J} &=&  m_1 ({\bf x}_1 \times {\bf v}_1)
\left ( 1 + \frac{1}{2} v_1^2 - \frac{1}{2} \frac{m_2}{r} \right )
- \frac{1}{2} \frac{m_1m_2}{r}(7\,{\bf x} \times {\bf
v}_{2} 
+ {\bf v}_2 \cdot {\bf n}\,{\bf x}_1 \times {\bf n} )
\nonumber \\
&&
+ {\bf S}_1 \left (1 + \frac{1}{2} v_1^2 + 3 \frac{m_2}{r} \right )
-\frac{1}{2} {\bf v}_1 \times ( {\bf v}_1 \times {\bf S}_1) 
\nonumber \\
&&
+ \frac{m_2}{r} \biggl ( 2{\bf n} - \frac{1}{2}
\frac {{\bf x}_1}{r} \biggr ) \times ({\bf n} \times {\bf S}_1)
 + (1 \rightleftharpoons 2 ) \,,
 \label{JPN}
\\
{\bf I} &=& m_1 {\bf x}_1 \left ( 1 + \frac{1}{2} v_1^2 - \frac{1}{2}
\frac{m_2}{r} \right ) +  \frac{1}{2} {\bf v}_1 \times {\bf S}_1 
 + (1 \rightleftharpoons 2 )\,,
 \label{IPN}
\eea
\label{EJPN}
\ees
where, to the 1PN order needed, the masses shown are the
total masses of each body,
given by
$m_1 + \frac{1}{2} \Omega_1$ and
$m_2 + \frac{1}{2} \Omega_2$.
The first thing to notice about these conserved quantities is the absence of
a spin-orbit contribution to the total energy.  This well-known result
is merely a consequence
of our choice of SSC.  Converting from our SSC
to the ``covariant'' SSC, for example, gives the standard spin-orbit term
(see Appendix \ref{app:ssc=1}).  These conserved quantities can also be
derived from global definitions of energy, momentum 
and angular momentum, as discussed
in Appendix \ref{app:EJ}.  It is straightforward to verify, by taking a time
derivative of Eqs. (\ref{EJPN}) and substituting the 1PN equations of motion
(\ref{a1PN}), (\ref{a1SO}) and spin precession (\ref{spinprecess}), that 
$E$, ${\bf J}$ and ${\dot {\bf I}}$ are constant to 1PN order.  

\subsection{The proper spin}
\label{sec:properspin}

The total angular momentum ${\bf J}$, Eq. (\ref{JPN}),
 has been written in a form
that appears to have
an orbital piece, plus 1PN corrections, a spin piece, plus 1PN
corrections, and a final, spin-orbit piece.  Although the split is somewhat
arbitrary, it is useful in that, if we identify the ``proper'' spin of each
body by the collection of ``spin'' terms in Eq. (\ref{JPN}), then
Eq. (\ref{spinprecess}) is equivalent to the standard spin-precession
equation (\ref{spinsummary}).   In addition, we found that, in our gauge,
there was a 2.5PN contribution to the spin precession [Eq. (\ref{spin2.5PN})],
but since that contribution was a total time derivative to 2.5PN order, it
could be absorbed into a redefinition of the spin.  We therefore define the
proper spin of each body to be
\bea
{\cal S}_1^i &\equiv&
  S_1^i \biggl (1 + \frac{1}{2} v_1^2 + 3 \frac{m_2}{r} \biggr )
  -\frac{1}{2} [{\bf v}_1 \times ( {\bf v}_1 \times {\bf S}_1)]^i 
   - S_1^i \stackrel{(3)}{{\cal I}^{jj}} + S_1^j \stackrel{(3)}{{\cal
  I}^{ij}} \,,
  \nonumber \\
{\bf {\cal S}}_2 &\equiv& (1 \rightleftharpoons 2 ) \,,
\label{properspin}
\eea
With this definition, the spins ${\bf {\cal S}}_A$ satisfy Eq.
(\ref{spinsummary}), with {\em no} 2.5PN contributions.  


\subsection{System center of mass and transformation to relative
coordinates}
\label{centermass}

Choosing our coordinates so that the ``center of mass'' quantity ${\bf I}$
vanishes, and defining 
the transformation from individual to relative coordinates 
to 1PN order by
\begin{eqnarray}
x_1^i &=& \frac{m_2}{m} x^i + \delta x^i \,,
\nonumber \\
x_2^i &=& -\frac{m_1}{m} x^i + \delta x^i \,,
\label{transform}
\end{eqnarray}
we obtain
\beq
\delta x^i = \frac{1}{2} \eta \frac{\delta m}{m} \left (v^2 - \frac{m}{r}
\right ) x^i 
- \frac{1}{2} \frac{\eta}{\delta m} \epsilon^{ijk} v^j ({\cal S}^k - \xi^k)
\,,
\label{transform2}
\eeq
where $\delta m \equiv m_1 -m_2$.
These transformations do not affect the Newtonian term in the acceleration, 
of course.  In the 1PN and spin-orbit terms they will only produce 2PN
effects, which we ignore.
The multipole moments
that appear in the 2.5PN terms in the equation of motion (\ref{a12.5PN})
must
also be converted to relative coordinates, keeping any PN and spin-orbit
corrections
generated by Eqs. (\ref{transform}); this is treated in Appendix
\ref{app:moments}.  In addition, in the 2.5PN terms,
multiple time derivatives of the multipole moments will generate
accelerations, for which the 1PN relative equations of motion including
spin-orbit terms must be
substituted; in explicitly 3.5PN terms, the Newtonian equation of
motion suffices.

Calculating the relative acceleration ${\bf a}={\bf a}_1-{\bf a}_2$
using Eqs. (\ref{a1summary0}) and (\ref{a1summary}), and converting to
relative coordinates, we obtain the equations of motion given by Eqs.
(\ref{eomsummary}), (\ref{eomsummary2}) and (\ref{arel2.5PN}).  

In terms of these relative coordinates, the energy and angular momentum of the
system to 1PN order including spin terms then take the form, 
\bes
\bea
E &=& \mu \left \{ \frac{1}{2}v^2 - \frac{m}{r}
+ \frac{3}{8} (1-3\eta)v^4 + \frac{1}{2} (3+\eta)v^2\frac{m}{r}
+\frac{1}{2}\eta\frac{m}{r} {\dot r}^2 +\frac{1}{2} \left ( \frac{m}{r}
\right )^2 \right \}
\,,
\\
{\bf J} &=&
\mu {\bf {\tilde L}}_{\rm N} \left \{ 1 + \frac{1}{2}(1-3\eta)v^2 
+(3+\eta)\frac{m}{r} \right \}
+ {\bf {\cal S}} + \frac{1}{2} \frac{\mu}{r} {\bf n} \times [ {\bf n} \times
(4 {\bf {\cal S}} + 3 \fourvec{\xi} )] \,,
\eea
\label{EJrelative}
\ees
where $\mu = \eta m$ is the reduced mass.
Notice that we do not keep the 2.5PN contribution to $\bf J$ arising from
the conversion from baryonic spin to proper spin using Eq.
(\ref{properspin}), since $E$ and ${\bf J}$ are only well defined up to 2PN
order.
In Sec.  \ref{compareflux} we will use these expressions
together with the 3.5PN equations of motion to compare $\dot E$ and $\dot
{\bf J}$ with the corresponding fluxes of radiation to infinity.

\section{3.5PN equations of motion}
\label{sec3.5pn}

\subsection{Equation of motion}

To obtain the 3.5PN contributions to the equations of motion including spin
terms, we take the 3.5PN fluid expressions shown in Appendix
\ref{app:fluid}, multiply by $\rho^*$, and integrate over body 1.  We follow
the same procedure as in Sec. \ref{sec2.5pnA}, expanding potentials about the
baryonic centers of mass of the bodies, keeping 
point-mass terms, internal self-energy terms,
and spinlike terms (terms linear in ${\bar x}{\bar v}$) and discarding
tidal-like terms. 
Many terms in Eq. (\ref{3.5eomfluid}) 
make only point-mass contributions, such as terms of the 
form $r^2 x^k (d^7 {\cal I}^{ik}/dt^7)$,  $x^j U (d^5 {\cal I}^{ij}/dt^5)$,
$v^2 v^i (d^4 {\cal I}^{kk}/dt^4)$, 
$UU^{,j} (d^3 {\cal I}^{ij}/dt^3)$, and so on, because they do not have 
the proper 
mix of $x$ and $v$ to generate a spin.  However terms such as 
$r^2 v^k (d^6 {\cal I}^{ik}/dt^6)$, $x^j {\dot U} (d^4 {\cal I}^{ij}/dt^4)$ or
$\Phi^{,i} (d^3 {\cal I}^{kk}/dt^3)$ will generate spins.  

Some of these terms generate contributions of the form of an integral that
is purely internal to body
1 multiplied by a multipole moment; examples include 
${\cal T}_1^{ij} x_1^k d^5 {\cal I}^{jk}/dt^5$, 
$\Omega_1^{jk} v_1^i d^4  {\cal I}^{jk}/dt^4$,
${\cal H}_1^{ji} v_1^k d^3 {\cal I}^{jk}/dt^3$.  
Virial relations must then be applied
to these internal integrals, to see if any spin terms arise.   
But in the virial relation (\ref{virial1})
involving ${\cal T}_1^{ij}$ and $\Omega_1^{ij}$,
spin terms occur one PN order higher, hence there will be no contributions
from these terms to the acceleration at 3.5PN order.  On the other hand, the
virial relation (\ref{virial2}) involving ${\cal H}^{ij}$ {\em does} have a
spin contribution at the same order, so those virial-induced terms must be
kept.

To lowest order,
the mass multipole moments ${\cal I}^{ij\dots}$ and their derivatives do not
contain spin terms, however the current moments ${\cal J}^{ij\dots}$ do contain
both point-mass and spin terms at lowest order.  

In addition, the combination of 1PN terms $ 4{\dot V}^i
+\frac{1}{2} {\ddot X}^{,i}$ in Eq. (\ref{fluidPN}),
will generate accelerations
whose 2.5PN terms will produce 3.5PN point-mass and spin terms that must be
included.
We must also re-express the 1PN spin-orbit terms of Eq. (\ref{a1SO}) in terms
of the proper spin of Eq. (\ref{properspin}); the 2.5PN contributions there
will generate 3.5PN terms in the equation of motion.  Finally, in the 2.5PN
accelerations of Eq. (\ref{a12.5PN}), we must include the 1PN corrections to
the multipole moments as well as the 1PN terms in the equations of motion
that are generated by the many time derivatives; 
these corrections will also contain spin contributions
(Appendix \ref{app:moments}).  

The result for the 3.5PN acceleration of body 1 is an expression
too lengthy to reproduce here.  The point mass terms reproduce Eq. (4.2) of
Paper II, apart from small differences resulting from our use of the gauge
transformation (\ref{deltax2.53.5}) 
to remove purely time dependent terms from the 2.5PN
and 3.5PN equations of motion.  After transforming to relative
coordinates and obtaining the relative acceleration, our point-mass
terms match the corresponding expressions of Paper II precisely.
The expression involving spins
is equally lengthy.  

Calculating $a_1^i - a_2^i$ using Eqs. (\ref{a1PN}),
(\ref{a1SO}), (\ref{a12.5PN}) and our 3.5PN terms,
substituting Eqs. (\ref{transform}) and the time-derivatives of the
multipole moments (\ref{relativemoments}), and expressing the spin results
in terms of the total spin ${\bf {\cal S}}$ and the spin quantity
$\fourvec{\xi}$, we obtain
the final relative equation of motion 
terms as given in Eqs. (\ref{eomsummary2}), (\ref{arel2.5PN}) 
and (\ref{aSORR}).

\subsection{Spin precession}

We now want to calculate the precession of the proper spin ${\bf {\cal S}}_1$
to 3.5PN order.  A time derivative of Eq. (\ref{properspin}) gives
\bea
{\dot {\cal S}}_1^i &=& {\dot S}_1^i 
\biggl (1 + v_1^2 + 3 \frac{m_2}{r} \biggr )
   -\frac{1}{2} v_1^i ( {\bf v}_1 \cdot {\dot {\bf S}}_1)
     \nonumber\\
&&
+ S_1^i \left ( 2 {\bf v}_1 \cdot {\bf a}_1 -3\frac{m_2 {\dot r}}{r^2} \right )
-\frac{1}{2} a_1^i ({\bf v}_1 \cdot {\bf S}_1)
-\frac{1}{2} v_1^i ({\bf a}_1 \cdot {\bf S}_1)
\nonumber \\
       && 
 - S_1^i \stackrel{(4)}{{\cal I}^{jj}} 
 + S_1^j \stackrel{(4)}{{\cal I}^{ij}}
 - {\dot S}_1^i \stackrel{(3)}{{\cal I}^{jj}} 
 + {\dot S}_1^j \stackrel{(3)}{{\cal I}^{ij}} \,.
\label{dotproperspin}
\eea
We repeat the method of Sec. \ref{sec2.5pnA} to determine the contributions
of 3.5PN fluid terms to the time derivative of the baryonic spin ${\bf S}_1$,  
by calculating $\epsilon^{ijk}\int_1 \rho^* {\bar x}^j a_{\rm 3.5PN}^k d^3x$.
Only terms in $a_{\rm 3.5PN}^k$ [Eq. (\ref{3.5eomfluid})] that have explicit $v$
dependence will contribute a spin term.   Notice that, as we have discussed,
the 2.5PN contribution to ${\dot S}_1^i$ cancels the relevant terms in the
last line of Eq. (\ref{dotproperspin}).  For ${\bf a}_1$, which appears in
the 1PN terms in Eq. (\ref{dotproperspin}), 
we must substitute the 2.5PN equations of motion; for ${\dot
S}_1^i$ in the final 2.5PN terms in Eq. (\ref{dotproperspin}) we must
substitute the 1PN precession equations; finally we must use Eq.
(\ref{properspin}) to convert from ${\bf S}_1$ back to the proper spin ${\bf
{\cal S}}_1$ to the appropriate order.  

The result is the 1PN spin precession of Eq. (\ref{spinsummary}), plus a
lengthy 3.5PN expression.
However, using the fact that, to lowest order ${\dot {\bf {\cal S}}}_1 = 0$,
together with the identities listed in Appendix \ref{app:timederiv}, it is
straightforward to show that our lengthy 3.5PN expression is in fact a
total time derivative, given by
\bea
({\dot {\bf {\cal S}}}_1)_{\rm 3.5PN} &=& \frac{\eta^2m^2}{5}\frac{d}{dt} \left (
\frac{\dot r}{r^2} \left \{ 
{\bf S}_1 \left [ (14-13\alpha)v^2-(10+15\alpha){\dot
r}^2 - \frac{2}{3}(27+11\alpha)\frac{m}{r} \right ]
\right .
\right .
\nonumber \\
&&
\left .
\left .
+{\bf n} {\bf S}_1 \cdot {\bf n}
\left [ 15(4-\alpha)v^2 - 25(4+3\alpha){\dot r}^2 - \frac{10}{3}(17+37\alpha)
\frac{m}{r} \right ]
-{\bf v} {\bf S}_1 \cdot {\bf v}(40+25\alpha)
\right \}
\right .
\nonumber \\
&&
\left .
-\frac{1}{r^2}  {\bf n} {\bf S}_1 \cdot {\bf v} 
\left [ (18-14\alpha)v^2 - 6(9+10\alpha){\dot r}^2 -
\frac{1}{3}(34+161\alpha)\frac{m}{r} \right ]
\right .
\nonumber \\
&&
\left .
-\frac{1}{r^2}  {\bf v} {\bf S}_1 \cdot {\bf n} 
\left [ (26-20\alpha)v^2-(66+45\alpha){\dot r}^2 
-\frac{5}{3}(50+59\alpha)\frac{m}{r} \right ] 
\right ) \,,
\label{Sdot3.5}
\eea
where $\alpha = m_2/m_1$.
As such, it can be moved to the left-hand-side and absorbed into a 
meaningless, 3.5PN
term in the redefinition of the spin.
As a result, we find, to little surprise, that radiation reaction makes {\em
no}
contribution to the precession of the spins [for a physical justification,
see remarks following Eq.
(\ref{spinrr})].

\subsection{Comparison with fluxes of energy and angular momentum}
\label{compareflux}

The fluxes of energy and angular momentum in gravitational waves from a binary
with spin-orbit interactions were derived by Kidder {\em et al.}
\cite{kww,kidder}, and are given by
\bea
\frac{dE}{dt} &=& {\dot E}_{\rm N}  + {\dot E}_{\rm SO} \,,
\nonumber \\
\frac{d{\bf J}}{dt} &=& {\dot {\bf J}}_{\rm N} + {\dot {\bf J}}_{\rm SO} \,,
\eea
where we include only the lowest-order ``Newtonian'' and 1PN spin-orbit
contributions.   After transforming from Kidder's $k_{\rm SSC}=1$ formulae,
Eqs. (3.25a), (3.25c), (3.28a) and (3.28c) of Ref. \cite{kidder},
to our $k_{\rm SSC}=1/2$ using
the transformation ${\bf x} \to {\bf x} - ({\bf v}\times \fourvec{\xi})/2m$,
we obtain for the fluxes,
\bes
\bea
{\dot E}_N &=& -\frac{8}{15} \frac{\eta^2 m^4}{r^4} (12v^2 -11{\dot r}^2) \,,
\label{ENflux}
\\
{\dot E}_{SO} &=&  -\frac{8}{15}\frac{\eta^2 m^3}{r^6} 
\left [ {\bf {\tilde L}}_{\rm N} \cdot {\bf {\cal S}} \left ( 27{\dot r}^2
- 37v^2-12\frac{m}{r} \right ) 
\right .
\nonumber \\
&& 
\left .
+ {\bf {\tilde L}}_{\rm N} \cdot \fourvec{\xi} \left (18{\dot r}^2
-19v^2 - 8\frac{m}{r} \right )
\right ] \,,
\label{ESOflux}
\\
{\dot {\bf J}}_N &=& -\frac{8}{5} \frac{\eta^2 m^3}{r^3} 
{\bf {\tilde L}}_{\rm N} \left ( 2v^2 - 3{\dot r}^2 - 2\frac{m}{r} \right )
\,,
\label{JNflux}
\\
{\dot {\bf J}}_{SO} &=& 
-\frac{4}{5}\frac{\eta^2m^2}{r^3} \left \{
\frac{2}{3}\frac{m}{r}(v^2-{\dot r}^2)({\bf {\cal S}}-\fourvec{\xi})
-\frac{1}{3}{\dot r}\frac{m}{r} {\bf n} \times ( 7 {\bf v}\times {\bf
{\cal S}} + 5 {\bf v}\times \fourvec{\xi})
\right .
\nonumber \\
&&
\left .
+\frac{m}{r} {\bf n}\times \left [
( {\bf n}\times {\bf {\cal S}}) 
\left ( 6{\dot r}^2 - \frac{17}{3}v^2 + 2\frac{m}{r} \right )
+( {\bf n}\times \fourvec{\xi})
\left ( 6{\dot r}^2 - 6v^2 + \frac{4}{3}\frac{m}{r} \right )
\right ]
\right .
\nonumber \\
&&
\left .
+{\dot r}{\bf v}\times\left [
( {\bf n}\times {\bf {\cal S}}) 
\left ( \frac{29}{3}\frac{m}{r} + 24v^2 - 30 {\dot r}^2 \right )
+ 5( {\bf n}\times \fourvec{\xi})
\left ( \frac{5}{3}\frac{m}{r} + 4v^2  -5 {\dot r}^2 \right )
\right ]
\right .
\nonumber \\
&&
\left .
+ {\bf v}\times\left [
{\bf v}\times {\bf {\cal S}}
\left ( 18 {\dot r}^2 - 12v^2 - \frac{23}{3}\frac{m}{r} \right )
+ {\bf v}\times \fourvec{\xi}
\left ( 15 {\dot r}^2 - \frac{29}{3}v^2 - 7\frac{m}{r} \right )
\right ]\right .
\nonumber \\
&&
\left .
+ \frac{1}{r^2} {\bf {\tilde L}}_{\rm N} {\bf {\tilde L}}_{\rm N} \cdot
\left [ 
{\bf {\cal S}}
\left ( 30{\dot r}^2 - 18 v^2 - \frac{92}{3}\frac{m}{r} \right )
+ \fourvec{\xi} 
\left (20{\dot r}^2 - 13v^2 - \frac{59}{3}\frac{m}{r} \right )
\right ]
\right \} \,.
\label{JSOflux}
\eea
\label{EJfluxes}
\ees

We now calculate the time derivative of the energy and angular momentum
expressions (\ref{EJrelative}), and substitute the 
equations of motion, including 1PN, spin-orbit, 2.5PN and 3.5PN spin orbit 
terms, along with the 1PN spin precession equations (recall there are no
3.5PN contributions to the spin precession).  
After recovering the fact that all 1PN point-mass
and spin-orbit contributions cancel,
leaving $E$ and ${\bf J}$ conserved to that order, we find that the changes
in $E$ and ${\bf J}$ due to 2.5PN and 3.5PN spin-orbit radiation reaction 
are obtained from the following expressions,
\bea
{\dot E} &=& \mu {\bf v} \cdot ({\bf a}_{\rm 2.5PN} + {\bf a}_{\rm
3.5PN-SO}) \,,
\nonumber \\
{\dot {\bf J}}  &=& \mu {\bf x} \times ({\bf a}_{\rm 2.5PN} + {\bf a}_{\rm
3.5PN-SO}) \,.
\eea
Initially, the results do not
match the flux expressions above.  However, by making use of the identities
listed in Appendix \ref{app:timederiv}, we can show that the difference between
the expressions in all cases is a total time derivative.  These can thus be
absorbed into meaningless 2.5PN and 3.5PN corrections to the definition of
total energy and angular momentum.  Thus we have established a proper energy
and angular momentum balance between the radiation flux and the evolution of
the orbit, including spin-orbit effects.

\section{Conclusions}

We have derived the equations of motion for binary systems of spinning
bodies from first principles, including the
effects of 
gravitational radiation reaction, and incorporating the contributions of
spin-orbit coupling at 3.5PN order.  We found that the spins themselves are
unaffected by radiation reaction.  The resulting equations of motion are
instantaneous, dynamical equations, and do not rely on assumptions of energy
balance, or orbital averaging.  They may be used to study the effects of
spin on the inspiral of compact binaries numerically.  
We have focussed attention on effects linear in the spins,
corresponding to spin-orbit coupling;  the effects of spin-spin
coupling can in principle also be calculated with our approach.  These
issues will be the subject of future work.

\acknowledgments

This work is supported in part by the National Science Foundation under
grant numbers PHY 00-96522 and PHY 03-53180, and by the National Aeronautics
and Space Administration under grant number NAG5-10186.  Achamveedu
Gopakumar made useful contributions to this work in its early phases.  
We are grateful to the  
Groupe Gravitation Relativiste et Cosmologie (GR$\varepsilon$CO) of the 
Institut d'Astrophysique de Paris for its hospitality during the academic
year 2003-04, where this work was completed.

\appendix

\section{Spin supplementary conditions}
\label{app:ssc}

When we deal with systems containing spinning bodies, the fact that the
bodies have finite size introduces an ambiguity in the definition of each
body's center of mass.  This has given rise to the concept of ``spin
supplementary condition'' (SSC) \cite{papapetrou1,papapetrou2}, 
which is a  condition whose role is to fix
the center of mass.  
One defines the antisymmetric tensor $S_A^{\mu\nu}$,
given by
\beq
S_A^{\mu\nu} \equiv 2 \int_A (x^{[\mu} - x_A^{[\mu} ) \tau^{\nu]0} d^3x \,,
\label{spin4tensor}
\eeq
where $\tau^{\alpha\beta}$ is 
the source stress-energy pseudo-tensor that appears
in the ``relaxed Einstein equations'' given by $\Box h^{ \alpha \beta } =
-16 \pi {\tau}^{ \alpha \beta }$, and is a combination of the stress-energy
tensor of matter and terms quadratic in the gravitational fields $h^{ \alpha
\beta }$ (see Sec.~II.A. of Paper I), and 
where $x_A^\mu$ is to be identified with the world line of the
center of mass of the body.  
The center of mass is then fixed by imposing the following 
condition:
\beq
S_A^{i0} - k_{\rm SSC} S_A^{ij}v_A^j =0 \,,
\eeq
where $k_{\rm SSC}$ typically has the values $1$, $1/2$ or $0$.  The value 
$k_{\rm SSC}=1$
corresponds to the so-called ``covariant'' SSC, $S_A^{\mu\nu}u_{A\nu} =0$.  

In this paper, our baryonic center-of-mass
definition $x_A^i = \int_A \rho^* x^i d^3x$ corresponds
to $k_{\rm SSC}=1/2$, which can be seen as follows.   
From Eq. (\ref{tau00}),
we note that, 
to the appropriate order,
$\tau^{00} = \rho^* (1 + \frac{1}{2}v^2 - \frac{1}{2} U)$.  Calculating
$S^{i0}$ following the methods outlined in Sec. \ref{sec2.5pn}, and
including the spin term generated by $\int_A \rho^* x^i v^2 d^3x$, we find
directly
that $S_A^{i0} = \frac{1}{2} S_A^{ij} v_A^j$, and thus that $k_{\rm
SSC}=1/2$.  

One can also show that the relationship between the centers of mass for each
value of $k_{\rm SSC}$ is given by
\beq
(x_A^i)^{(k^\prime)} = (x_A^i)^{(k)} + \frac{k-k^\prime}{m_A}
S_A^{ij}(v_A^j)^{(k)}
\,.
\eeq
For further discussion of the role of the SSC in PN calculations, see
\cite{barkeroconrev,kidder,barkeroconssc}.

\section{Equations of motion with spin in the covariant SSC}
\label{app:ssc=1}

We now transform our key equations to a form in which the centers of mass of
the bodies are defined by the covariant spin supplementary condition,
$k_{\rm SSC}=1$.  That transformation is given by 
\beq
{\bf x}_1^{(1/2)} = {\bf x}_1 + \frac{1}{2m_1} {\bf v}_1 \times {\bf
S}_1 \,,
\eeq
where the variables on the right-hand-side are in terms of $k_{\rm SSC}=1$,
and where ${\bf S}_1$ is the baryonic spin.  For the relative coordinate,
the transformation, to 1PN order, is
\beq
{\bf x}^{(1/2)} = {\bf x} + \frac{1}{2m} {\bf v} \times \fourvec{\xi}
\,,
\eeq
where $\fourvec{\xi}$ is again the baryonic spin variable.
Converting from the baryonic spin to proper spin and keeping only the 2.5PN
correction from Eq. (\ref{properspin}) (the PN corrections will not be
relevant for this purpose), we have 
\beq
{\bf x}^{(1/2)} = {\bf x} + \frac{1}{2m} {\bf v} \times \fourvec{\xi}
-\frac{1}{2m} {\bf v} \times \fourvec{\xi} \stackrel{(3)}{{\cal I}^{kk}}
+\frac{1}{2m} {\bf v} \times (\xi^j \stackrel{(3)}{{\cal I}^{jk}}) {\bf e}_k
\,,
\eeq
where $\fourvec{\xi}$ now represents the proper spin variable.
We substitute this transformation into the Newtonian, PN, spin-orbit and 2.5PN
terms in the equations of motion, Eqs. (\ref{eomsummary}),
(\ref{eomsummary2}) and (\ref{aSORR}), keeping only PN spin-orbit terms,
2.5PN terms, and 3.5PN spin-orbit terms.  Where accelerations arise, {\em
eg.} in transforming the velocities in the spin-orbit terms or in the 2.5PN
terms, we must
employ suitably accurate expressions, in order to generate all appropriate
3.5PN spin-orbit terms.  The result is to change the form only of the
1PN spin-orbit and
3.5PN spin-orbit terms, which now become 
\bes
\bea
{\bf a}_{\rm SO}  &=& \frac{1}{r^3} \biggl \{
6 \frac{\bf n}{r} {\bf {\tilde L}}_{\rm N} \cdot
\left ( {\bf {\cal S}} + \fourvec{\xi} \right )
- {\bf v} \times
\left ( 4{\bf {\cal S}} + 3\fourvec{\xi} \right )
+3 \dot r {\bf n} \times
\left ( 2{\bf {\cal S}} + \fourvec{\xi} \right )
\biggr \} \,,
\label{aPNSOssc=1}
\\
{\bf a}_{\rm 3.5PN-SO} &=& 
-\frac{\eta m}{5r^4} 
\biggl \{
\frac{{\dot r}{\bf n}}{r}
\left [ \left ( 120v^2+280{\dot r}^2+453\frac{m}{r} \right )
{\bf {\tilde L}}_{\rm N} \cdot {\bf {\cal S}}
\right .
\nonumber \\
&&
\left .
+ \left ( 120v^2+280{\dot r}^2+458\frac{m}{r} \right )
{\bf {\tilde L}}_{\rm N} \cdot \fourvec{\xi}
\right ]
\nonumber \\
&&
+
\frac{\bf v}{r}
\left [ \left ( 87v^2-675{\dot r}^2-\frac{901}{3}\frac{m}{r}
\right )
{\bf {\tilde L}}_{\rm N} \cdot {\bf {\cal S}}
+ 4\left ( 18v^2-150{\dot r}^2 - 66\frac{m}{r} \right )
{\bf {\tilde L}}_{\rm N} \cdot \fourvec{\xi}
\right ]
\nonumber \\
&&
- \frac{2}{3}{\dot r}{\bf v} \times {\bf {\cal S}}
\left ( 48v^2 + 15{\dot r}^2+364\frac{m}{r} \right )
+ \frac{1}{3}{\dot r}{\bf v} \times \fourvec{\xi}
\left ( 291v^2 -705{\dot r}^2-772\frac{m}{r} \right )
\nonumber \\
&&
+\frac{1}{2}{\bf n}\times {\bf {\cal S}}
\left ( 31v^4-260v^2{\dot r}^2+245{\dot r}^4
-\frac{689}{3}v^2\frac{m}{r} + 537{\dot r}^2\frac{m}{r}
+\frac{4}{3}\frac{m^2}{r^2} \right )
\nonumber \\
&&
+\frac{1}{2}{\bf n}\times \fourvec{\xi}
\left ( 115v^4-1130v^2{\dot r}^2+1295{\dot r}^4
-\frac{869}{3}v^2\frac{m}{r} + 849{\dot r}^2\frac{m}{r}
+ \frac{44}{3}\frac{m^2}{r^2} \right )
\biggr \} \,.
\label{aSORRssc=1}
\eea
\label{aSOssc=1}
\ees
The spin precession equations are not affected by this SSC transformation,
since we are working to linear order in spins.
The energy and angular momentum in this SSC take the form
\bes
\bea
E &=& \mu \left \{ \frac{1}{2}v^2 - \frac{m}{r}
\right .
\nonumber\\
&&
\left .
+ \frac{3}{8} (1-3\eta)v^4 + \frac{1}{2} (3+\eta)v^2\frac{m}{r}
+\frac{1}{2}\eta\frac{m}{r} {\dot r}^2 +\frac{1}{2} \left ( \frac{m}{r}
\right )^2 
\right .
\nonumber\\
&&
\left .
+ \frac{1}{r^3} {\bf {\tilde L}}_{\rm N} \cdot \fourvec{\xi} 
\right \}
\,,
\\
{\bf J} &=&
\mu {\bf {\tilde L}}_{\rm N} \left \{ 1 + \frac{1}{2}(1-3\eta)v^2
+(3+\eta)\frac{m}{r} \right \}+ {\bf {\cal S}} 
\nonumber\\
&&
+ \frac{1}{2} \frac{\mu}{m} \left \{ \frac{m}{r} 
{\bf n} \times [ {\bf n} \times (4 {\bf {\cal S}} + 2 \fourvec{\xi} )] \,.
- {\bf v} \times [ {\bf v} \times \fourvec{\xi} ] \right \}
\,.
\eea
\label{EJrelative1}
\ees
The spin-orbit contributions to the energy and angular momentum fluxes, from
Eqs. (3.25c) and (3.28c) of \cite{kidder} are given by
\bes
\bea
{\dot E}_{SO} &=&  -\frac{8}{15}\frac{\eta^2 m^3}{r^6} 
\left [ {\bf {\tilde L}}_{\rm N} \cdot {\bf {\cal S}} \left ( 27{\dot r}^2
- 37v^2-12\frac{m}{r} \right ) 
\right .
\nonumber \\
&& 
\left .
+ {\bf {\tilde L}}_{\rm N} \cdot \fourvec{\xi} \left (51{\dot r}^2
-43v^2 +4\frac{m}{r} \right )
\right ] \,,
\label{ESOflux1}
\\
{\dot {\bf J}}_{SO} &=& 
-\frac{4}{5}\frac{\eta^2m^2}{r^3} \left \{
\frac{2}{3}\frac{m}{r}(v^2-{\dot r}^2)({\bf {\cal S}}-\fourvec{\xi})
-\frac{1}{3}{\dot r}\frac{m}{r} {\bf n} \times ( 7 {\bf v}\times {\bf
{\cal S}} + 5 {\bf v}\times \fourvec{\xi})
\right .
\nonumber \\
&&
\left .
+\frac{m}{r} {\bf n}\times \left [
( {\bf n}\times {\bf {\cal S}}) 
\left ( 6{\dot r}^2 - \frac{17}{3}v^2 + 2\frac{m}{r} \right )
+( {\bf n}\times \fourvec{\xi})
\left ( 9{\dot r}^2 - 8v^2 - \frac{2}{3}\frac{m}{r} \right )
\right ]
\right .
\nonumber \\
&&
\left .
+{\dot r}{\bf v}\times\left [
( {\bf n}\times {\bf {\cal S}}) 
\left ( \frac{29}{3}\frac{m}{r} + 24v^2 - 30 {\dot r}^2 \right )
+ 5( {\bf n}\times \fourvec{\xi})
\left ( \frac{5}{3}\frac{m}{r} + 4v^2  -5 {\dot r}^2 \right )
\right ]
\right .
\nonumber \\
&&
\left .
+ {\bf v}\times\left [
{\bf v}\times {\bf {\cal S}}
\left ( 18 {\dot r}^2 - 12v^2 - \frac{23}{3}\frac{m}{r} \right )
+ {\bf v}\times \fourvec{\xi}
\left ( 18 {\dot r}^2 - \frac{35}{3}v^2 - 9\frac{m}{r} \right )
\right ]\right .
\nonumber \\
&&
\left .
+ \frac{1}{r^2} {\bf {\tilde L}}_{\rm N} {\bf {\tilde L}}_{\rm N} \cdot
\left [ 
{\bf {\cal S}}
\left ( 30{\dot r}^2 - 18 v^2 - \frac{92}{3}\frac{m}{r} \right )
+ \fourvec{\xi} 
\left (35{\dot r}^2 - 19v^2 - \frac{71}{3}\frac{m}{r} \right )
\right ]
\right \} \,.
\label{JSOflux1}
\eea
\label{EJfluxes1}
\ees
By following the same steps as in Sec. \ref{compareflux}, calculating the
time derivative of $E$ and ${\bf J}$, substituting the relevant contributions
from the equations of motion and spin precession,
and extracting total time derivatives from the result (using $k_{SSC}=1$
versions of the equations in Appendix \ref{app:timederiv}),
we verify that the losses of orbital energy and angular momentum
are completely equivalent to the fluxes above.

\section{Multipole moments}
\label{app:moments}

The multipole moments that appear in the radiation-reaction terms in Eqs.
(\ref{fluid2.5PN}) and (\ref{3.5eomfluid}) 
are defined by the general expressions (\ref{sourcemoments}).  
Because the quadrupole moment ${\cal I}^{ij}$ appears in the 2.5PN
terms, it
will be needed to 1PN order, while the remaining moments will be needed to
only the lowest, ``Newtonian'' order.  Substituting the relevant expressions
for $\tau^{\alpha\beta}$ from Eqs. (\ref{taumunu}), and carrying out the
standard procedure as in Appendix \ref{app:EJ}, we obtain
\bea
{\cal I}^{ij} &=& m_1 x_1^{ij} \left ( 1+ \frac{1}{2}v_1^2
\right )
-  \frac{m_1m_2}{r}
\left ( \frac{1}{2} x_1^{ij} - \frac{7}{4}  r^2
\delta^{ij} \right ) + x_1^{(i} {\cal S}_1^{j)k} v_1^k 
\nonumber \\
&& + (1 \rightleftharpoons 2)
\,,
\nonumber \\
{\cal I}^{ijk\dots} &=&  m_1 x_1^{ijk\dots} +  (1 \rightleftharpoons 2)\,,
\nonumber \\
{\cal J}^{ij} &=& \epsilon^{iab} \left ( m_1 v_1^bx_1^{aj} 
+ x_1^{(a} {\cal S}_1^{j)b} \right ) +  (1 \rightleftharpoons 2) \,,
\nonumber \\
{\cal J}^{ijk} &=& \epsilon^{iab}  \left ( m_1 v_1^bx_1^{ajk} 
+ \frac{3}{2} x_1^{(jk} {\cal S}_1^{a)b} \right ) +  (1 \rightleftharpoons 2) \,,
\nonumber \\
{\cal M}^{ijkl} &=&  m_1 v_1^{ij} x_1^{kl}
- \frac{1}{2}  \frac{m_1m_2}{r} n^{ij} x_1^{kl}
\nonumber \\
&&
+ \frac{1}{12}  m_1m_2 r
( n^{ijkl} -n^{ij} \delta^{kl}
-n^{kl} \delta^{ij}
+ n^{i(k} \delta^{l)j}
+ n^{j(k} \delta^{l)i} - 2 \delta^{i(k} \delta^{l)j}
+ 2 \delta^{ij}  \delta^{kl} ) 
\nonumber \\
&&
+ 2  x_1^{(k} {\cal S}_1^{l)(i} v_1^{j)} +  (1 \rightleftharpoons 2) \,,
\eea
where parentheses around indices denote symmetrization.
Note that higher-order moments, such as ${\cal J}^{qjkl}$ or ${\cal
M}^{kkjji}$ appear only in 3.5PN terms that were transformed away in
Appendix \ref{app:fluid}, so their explicit forms are not needed.

Converting to relative coordinates, using the 1PN correct
transformation in the leading term of ${\cal I}^{ij}$, we obtain
\begin{eqnarray}
{\cal I}^{ij} &=& \eta m x^{ij} \left ( 1+ \frac{1}{2}(1-3\eta)v^2 -
\frac{1}{2}(1-2\eta)\frac{m}{r} \right )
+ \frac{7}{2} \eta m^2 r \delta^{ij} 
+\eta x^{(i} \xi^{j)k} v^k \,,
\nonumber \\
{\cal I}^{ijk} &=& - \eta \delta m x^{ijk}  \,,
\nonumber \\
{\cal I}^{ijkl} &=& \eta m (1-3\eta)  x^{ijkl} \,,
\nonumber \\
{\cal J}^{ij} &=& -\eta \delta m {\tilde L}_{\rm N}^i x^j - \eta \epsilon^{iab}
x^{(j} \Delta^{a)b} \,,
\nonumber \\
{\cal J}^{ijk} &=&  \eta m (1-3\eta)  {\tilde L}_{\rm N}^i x^{jk} 
+\frac{3}{2} \eta \epsilon^{iab} x^{(jk} \xi^{a)b} \,,
\nonumber \\
{\cal M}^{ijkl} &=& \eta m (1-3\eta) \left (v^{ij} -
\frac{1}{3} \frac{m}{r} n^{ij} \right ) x^{kl}
\nonumber \\
&&
- \frac{1}{6}  \eta m^2 r
(  n^{ij} \delta^{kl}
+n^{kl} \delta^{ij} - n^{i(k} \delta^{l)j}
- n^{j(k} \delta^{l)i} + 2 \delta^{i(k} \delta^{l)j}
- 2 \delta^{ij}  \delta^{kl} ) 
\nonumber \\
&&
+ 2 \eta x^{(k} \xi^{l)(i} v^{j)}\,,
\label{relativemoments}
\end{eqnarray}
where ${\bf {\tilde L}}_{\rm N} \equiv {\bf x} \times {\bf v}$ 
is the orbital angular
momentum per unit mass, $\xi^{ij} = (m_2/m_1) {\cal S}_1^{ij} 
+ (m_1/m_2) {\cal S}_2^{ij}$
and $\Delta^{ij} \equiv m ({\cal S}_2^{ij}/m_2-{\cal S}_1^{ij}/m_1) 
= (m/\delta m)(\xi^{ij}-{\cal S}^{ij})$.

Time derivatives of the moments may be calculated using the relative
equations
of motion in place of ${\ddot x}^i$; 1PN equations including spin
contributions must be used in the
leading term in ${\cal I}^{ij}$, while Newtonian equations are
sufficient for the remaining terms.  Since spin effects in the moments and
their time derivatives are
already at 1PN order, the spins themselves may be treated as constants.

\section{3.5PN terms in the fluid equations of motion}
\label{app:fluid}

We start with the 2.5PN and 3.5PN
accelerations terms in the fluid equations of motion, written in terms of
the conserved density $\rho^*$, Eqs. (2.24c) and
(2.24d) of Paper II.  
We then make a coordinate transformation at 2.5PN and 3.5PN order to
eliminate the purely time dependent terms.  These terms cancel when we
compute the
relative acceleration, but 
contribute to the gravitational-radiation induced recoil of the system.
In other words, we choose coordinates that are fixed with respect to the
recoiling center of mass of the system.
This will
simplify the transformation between the coordinates ${\bf x}_1$ and 
${\bf x}_2$ of the individual bodies 
and the relative coordinate ${\bf x}$, Eq. (\ref{transform}),
and will simplify 
our analysis of angular momentum (which can depend on
the choice of center of mass)
The required transformation is
\beq
x^i \to x^i + \delta x^i_{2.5PN} + \delta x^i_{3.5PN} \,,
\eeq
where
\bes
\bea
\delta x^i_{2.5PN} &=& \frac{2}{15}  \stackrel{(3)}{{\cal I}^{ijj}}
 - \frac{2}{3} \epsilon^{qij} \stackrel{(2)}{{\cal J}^{qj}} \,,
\label{deltax2.5}
\\
\delta x^i_{3.5PN} &=& \frac{23}{4200} \stackrel{(5)}{{\cal I}^{ijjkk}}
-\frac{2}{75} \epsilon^{qij}\stackrel{(4)}{{\cal J}^{qjkk}}
-\frac{1}{30} \stackrel{(3)}{{\cal M}^{kkjji}} \,.
\label{deltax3.5}
\eea
\label{deltax2.53.5}
\ees
However, because of the velocity dependence in the PN terms in the
equations of motion, 
the change in velocity at 2.5PN order, $\delta v^i_{2.5PN} =
\delta {\dot x}^i_{2.5PN}$ will induce additional
3.5PN contributions in the equations of motion given by
\bea
\delta a^i_{3.5PN}  &=& \delta v^j_{2.5PN} (2v^j U^{,i} + v^i  U^{,j}
- {\dot U}\delta^{ij} + V^{j,i} + {\dot X}^{,ij} )
\nonumber \\
&& + \frac{1}{2}\delta {\dot v}^j_{2.5PN} ( X^{,ij} - 8U\delta^{ij} ) \,.
\eea
When we add these changes to 
the 3.5PN terms in Eq. (2.4d) of Paper II, we obtain the final 
3.5PN contributions to the fluid equations of motion, 
given by 
\bea
a_{3.5PN}^i &=&
\frac{1}{210} (13r^2x^k \delta^{ij}-4r^2x^i \delta^{jk} - x^ix^jx^k) 
	 \stackrel{(7)}{{\cal I}^{jk}}
\nonumber 
\eea
\bea
&&
+\frac{1}{30} ( 10r^2v^k \delta^{ij} + 4(v \cdot x)x^k \delta^{ij}
	-r^2v^i \delta^{jk} -4x^ix^jv^k + 3x^jx^kv^i ) 
	\stackrel{(6)}{{\cal I}^{jk}}
\nonumber 
\eea
\bea
&&
+\frac{1}{15} ( 10(v \cdot x)v^k \delta^{ij} - x^kv^2 \delta^{ij}
	+2(v \cdot x)v^i \delta^{jk} +2x^iv^2 \delta^{jk}
	-5x^iv^jv^k + 4x^jv^iv^k )
	\stackrel{(5)}{{\cal I}^{jk}}
\nonumber 
\eea
\bea
&&
+\frac{1}{15} (5r^2U^{,j} - 35x^jU + 59X^{,j} ) 
\stackrel{(5)}{{\cal I}^{ij}}
+\frac{1}{30} (5r^2U^{,i} + 30x^iU - 34X^{,i} - 6x^jX^{,ij})
	\stackrel{(5)}{{\cal I}^{kk}}
\nonumber 
\eea
\bea
&&
+\frac{1}{90} (15x^jx^kU^{,i} - 6x^jX^{,ik} - 30x^iX^{,jk}
	-15r^2X^{,ijk} - 4Y^{,ijk} + 5x^lY^{,ijkl} )
 	\stackrel{(5)}{{\cal I}^{jk}}
\nonumber 
\eea
\bea
&&
+\frac{1}{3}v^i ( v^2 \delta^{jk} - v^jv^k) \stackrel{(4)}{{\cal I}^{jk}}
-\frac{1}{3} (2x^j{\dot U} + 22V^j - 14 {\dot X}^{,j} - 12v^kX^{,jk})
	\stackrel{(4)}{{\cal I}^{ij}}
\nonumber 
\eea
\bea
&&
+\frac{1}{18} (24x^jv^kU^{,i}+12v^ix^jU^{,k} + 12x^jV^{k,i}
	+54v^iX^{,jk} -72v^jX^{,ik} 
\nonumber 
\eea
\bea
&&
	+ 12x^j {\dot X}^{,ik}
	+60X^{j,ki} - 72X^{i,jk} - 5{\dot Y}^{,ijk} )
	\stackrel{(4)}{{\cal I}^{jk}}
	-\frac{1}{3} (6v^iU-8V^i-2{\dot X}^{,i}) 
	\stackrel{(4)}{{\cal I}^{kk}}
\nonumber 
\eea
\bea
&&
+(2v^2U^{,j}- 8UU^{,j} -8v^kV^{k,j}+3\Phi_1^{,j} -2\Phi_2^{,j}
	+{\ddot X}^{,j} ) \stackrel{(3)}{{\cal I}^{ij}}
\nonumber 
\eea
\bea
&&
+\frac{1}{3} (2v^2U^{,i} - 8v^iv^jU^{,j} - 24UU^{,i} -6v^i{\dot U}
	+8 {\dot V}^i + 16v^jV^{[i,j]} +3\Phi_1^{,i} -6\Phi_2^{,i}
	+{\ddot X}^{,i} ) \stackrel{(3)}{{\cal I}^{kk}}
\nonumber 
\eea
\bea
&&
+\frac{1}{6} (48v^jV^{k,i} -12v^jv^kU^{,i} - 6v^2X^{,ijk}
	+24v^iv^lX^{,jkl} 
	-24{\dot X}^{i,jk}
	-9X_1^{,ijk} 
	+ 6X_2^{,ijk} 
\nonumber 
\eea
\bea
&&
	+18v^i{\dot X}^{,jk} 
	- 48v^lX^{[i,l]jk} 
	+24UX^{,ijk} +24U^{,i}X^{,jk}
	- 18\Phi_1^{jk,i} 
	+6\Sigma^{,i}(X^{,jk}) - {\ddot Y}^{,ijk} )
	\stackrel{(3)}{{\cal I}^{jk}}
\nonumber 
\eea
\bea
&&
+\frac{1}{630} (10x^ix^j \delta^{kl} - 25 x^kx^l \delta^{ij}
	- 9r^2 \delta^{ij}\delta^{kl} )  \stackrel{(7)}{{\cal I}^{jkl}}
\nonumber 
\eea
\bea
&&
+\frac{1}{45} (4x^{[i}v^{j]} \delta^{kl} - 2(v \cdot x)
	\delta^{ij}\delta^{kl} -10x^kv^l \delta^{ij} )
	\stackrel{(6)}{{\cal I}^{jkl}}
\nonumber 
\eea
\bea
&&
-\frac{1}{45} (v^2 \delta^{ij}\delta^{kl} + 6v^iv^j \delta^{kl}
	+ 5 v^kv^l \delta^{ij} ) \stackrel{(5)}{{\cal I}^{jkl}}
-\frac{1}{45} (4U \delta^{jk} +10x^jU^{,k}-5X^{,jk} )
	\stackrel{(5)}{{\cal I}^{ijk}}
\nonumber 
\eea
\bea
&&
+\frac{1}{45} (7X^{,ij} - 10x^jU^{,i} )\stackrel{(5)}{{\cal I}^{jkk}}
+\frac{1}{54} (6x^jX^{,ikl} - Y^{,ijkl} ) \stackrel{(5)}{{\cal I}^{jkl}}
\nonumber 
\eea
\bea
&&
+\frac{4}{45} ({\dot U}\delta^{ij}-v^iU^{,j}-2v^jU^{,i} -V^{j,i}
	-{\dot X}^{,ij} ) \stackrel{(4)}{{\cal I}^{jkk}}
\nonumber 
\eea
\bea
&&
+\frac{1}{45} (3r^2 \epsilon^{qik} + 2x^ix^j \epsilon^{qjk}
	+4x^jx^k \epsilon^{qij} ) \stackrel{(6)}{{\cal J}^{qk}}
-\frac{16}{45} x^jv^k \epsilon^{qjk} \stackrel{(5)}{{\cal J}^{qi}}
\nonumber 
\eea
\bea
&&
+\frac{2}{45} (2(v \cdot x) \epsilon^{qik} -2x^iv^j \epsilon^{qjk}
	+ 5x^jv^i \epsilon^{qjk} + 12x^jv^k \epsilon^{qij}
	+4x^kv^j \epsilon^{qij} )\stackrel{(5)}{{\cal J}^{qk}}
\nonumber 
\eea
\bea
&&
-\frac{1}{9} (2U \epsilon^{qik} +2x^j U^{,i} \epsilon^{qjk}
	-4 x^jU^{,k} \epsilon^{qij} 
	+X^{,ij} \epsilon^{qjk}
	-4X^{,jk} \epsilon^{qij}
	-4x^lX^{,ijk} \epsilon^{qlj} ) \stackrel{(4)}{{\cal J}^{qk}}
\nonumber 
\eea
\bea
&&
+\frac{2}{9} (4v^jv^k \epsilon^{qij} - v^2 \epsilon^{qik} )
	\stackrel{(4)}{{\cal J}^{qk}}
-\frac{4}{9} x^j U^{,k} \epsilon^{qjk} \stackrel{(4)}{{\cal J}^{qi}}
+\frac{2}{9} {\dot U} \epsilon^{qik}\stackrel{(3)}{{\cal J}^{qk}}
\nonumber 
\eea
\bea
&&
-\frac{2}{9} (v^iU^{,j}+2v^jU^{,i}+V^{j,i}+{\dot X}^{,ij} )
	\epsilon^{qjk}\stackrel{(3)}{{\cal J}^{qk}}
-\frac{1}{840} x^i \stackrel{(7)}{{\cal I}^{jjkk}}
\nonumber 
\eea
\bea
&&
+\frac{1}{35} x^j \stackrel{(7)}{{\cal I}^{ijkk}}
+\frac{1}{40} v^i \stackrel{(6)}{{\cal I}^{jjkk}}
+\frac{1}{24} U^{,i} \stackrel{(5)}{{\cal I}^{jjkk}}
-\frac{1}{30} x^j \epsilon^{qij} \stackrel{(6)}{{\cal J}^{qkk}}
-\frac{1}{15} x^j \epsilon^{qik} \stackrel{(6)}{{\cal J}^{qjk}}
\nonumber 
\eea
\bea
&&
+\frac{1}{15} v^j (\epsilon^{qjk}\stackrel{(5)}{{\cal J}^{qik}}
	-\epsilon^{qik} \stackrel{(5)}{{\cal J}^{qjk}} 
	-\epsilon^{qij}\stackrel{(5)}{{\cal J}^{qkk}} )
-\frac{1}{30} x^i \stackrel{(5)}{{\cal M}^{kkjj}} 
-\frac{1}{15} x^j \stackrel{(5)}{{\cal M}^{kkij}} 
\nonumber 
\eea
\bea
&&
-\frac{1}{6} v^i \stackrel{(4)}{{\cal M}^{kkjj}} 
+\frac{2}{3} v^j \stackrel{(4)}{{\cal M}^{ijkk}}
+\frac{1}{6}U^{,i}\stackrel{(3)}{{\cal M}^{jjkk}}
+\frac{2}{3} U^{,j} \stackrel{(3)}{{\cal M}^{ijkk}}
-\frac{1}{3} X^{,ijk} \stackrel{(3)}{{\cal M}^{jkll}} \,,
\label{3.5eomfluid}
\end{eqnarray}
where the relevant potentials are given in Eqs. (\ref{potentials}) and the
multipole moments are given in Appendix \ref{app:moments}.

\section{Virial relations}
\label{app:virial}

Virial relations are statements about the internal structure of bound bodies
that may be used to simplify the equations of motion.  They assume that the
body is either stationary or periodic over a suitable timescale.  The
simplest such virial relation, used in classical mechanics, states that $\ddot
I/2 = 2T + \Omega$, where $I = \int \rho^* |{\bar x}|^2 d^3x$ is 
the scalar moment of inertia of the
body, $T$ is the internal kinetic energy, and $\Omega$ is the internal
gravitational potential energy.  For a stationary body, or when averaged
over several internal timescales, we can set $\ddot I =0$, and hence $2T +
\Omega = 0$.  Here we generalize this to a variety of tensorial virial
relations, including spin effects and effects at 2.5PN order.  

We define the tensorial moment of inertia of
body A by 
\beq
I_A^{ij} \equiv \int_A \rho^* {\bar x}^i{\bar x}^j d^3x \,.
\label{momentinertia}
\eeq
Taking a time derivative and setting it
to zero yields 
${\dot I}_A^{ij} = 2\int_A \rho^* {\bar x}^{(i}{\bar v}^{j)}d^3x = 0$.  But from
this and Eq. (\ref{spintensor}), we can conclude that 
\beq
\int_A {\bar x}^i{\bar v}^j d^3x = \frac{1}{2} S_A^{ij} \,.
\eeq
A second time derivative gives ${\ddot I}_A^{ij} = 2\int_A \rho^* {\bar
v}^{i}{\bar v}^{j}d^3x + 2\int_A \rho^* {\bar x}^{(i}{a}^{j)}d^3x$.  
For $a^j$,
we substitute
only the Newtonian, post-Newtonian and  2.5PN terms from Eqs. (\ref{fluid}) and 
(\ref{fluidterms}).  In the equations of motion, these virial relations will
be needed only to simplify PN terms, so for this 
application, only the Newtonian and 2.5PN terms
will be needed.  On the other hand, in the expression $E=\int_{\cal M}
\tau^{00} d^3x$ for the conserved total
energy to PN order, the virial relation will be needed to simplify the
Newtonian contribution, and thus the 1PN spin contributions to the virial
relation will be needed.
Expanding potentials about the center of mass of each
body as described in Sec. (\ref{sec2.5pnA}), keeping PN and 2.5PN 
spin terms and 2.5PN
``internal'' terms, discarding terms that vanish as $R^2$ and
higher as the body's size tends to zero, and setting time derivatives of the
moment of inertia tensor to zero, we obtain the virial relation
for body 1:
\bea
2T_1^{ij} + \Omega_1^{ij} &=&
\frac{m_2}{r^2} \left ( 2n^k S_1^{k(i} (v_1^{j)} - v_2^{j)})
+n^{(i} S_1^{j)k} (v_1^k - 2v_2^k) \right )
\nonumber\\
&& + S_1^{k(i} \stackrel{(4)}{{\cal I}^{j)k}}
-\frac{1}{3} \Omega_1^{ij} \stackrel{(3)}{{\cal I}^{kk}}
-3 \Omega_1^{ijkl}  \stackrel{(3)}{{\cal I}^{kl}} \,,
\label{virial1}
\eea
The relevant internal quantities used here are defined below.
A third derivative of ${I}_A^{ij}$ gives $\stackrel{...}I_A^{ij} =
2 \int_A \rho^* (3{\bar v}^{(i} a^{j)} + {\bar x}^{(i} {\dot a}^{j)})d^3x$.
The same procedure yields, for body 1,
\bea
2{\cal H}_1^{(ij)} -\frac{3}{2}{\cal K}_1^{ij} &=& -\frac{3}{2}  \frac{m_2}{r^3}S_1^{k(i}n^{j)k}
\nonumber \\
&& + \frac{1}{5} S_1^{k(i} \stackrel{(5)}{{\cal I}^{j)k}}
-3 \frac{m_2}{r^3} n^kn^{(i}(S_1^{j)l}-\frac{5}{2}S_1^{j)m}n^mn^l)
\stackrel{(3)}{{\cal I}^{kl}}
\nonumber \\
&&
+ 2 \Omega_1^{k(i}\stackrel{(4)}{{\cal I}^{j)k}}
-\frac{1}{6}\Omega_1^{ij}\stackrel{(4)}{{\cal I}^{kk}}
-\frac{3}{2}\Omega_1^{ijkl}\stackrel{(4)}{{\cal I}^{kl}}
\nonumber \\
&&
-\frac{1}{2} ( 12{\cal K}_1^{(ij)kl} + 6{\cal K}_1^{klij} - 15 {{\cal K}_1^*}^{ijkl} )
\stackrel{(3)}{{\cal I}^{kl}} \,,
\label{virial2}
\eea
where we have used the lowest-order virial relation from Eqs.
(\ref{virial1}) and (\ref{virial2}), $2T_1^{ij} + \Omega_1^{ij} = 0$, 
and $2{\cal H}_1^{(ij)} -\frac{3}{2}{\cal K}_1^{ij} = -\frac{3}{2}
\frac{m_2}{r^3}S_1^{k(i}n^{j)k}$ to simplify some
of the 2.5PN order terms in Eq. (\ref{virial2}).  
The relevant quantities used in
these virial relations are defined by
\bea
{\cal T}_A^{ij}  &\equiv&\frac{1}{2} \int_A \rho^* {\bar v}^i{\bar v}^jd^3x
\,,
\nonumber \\
\Omega_A^{ij} &\equiv& -\frac{1}{2} \int_A \int_A \rho^* {\rho^*}^\prime
\frac{(x-x^\prime)^i(x-x^\prime)^j}{|{\bf x} - {\bf x}^\prime |^3} d^3x
d^3x^\prime \,,
\nonumber \\
\Omega_A^{ijkl} &\equiv& -\frac{1}{2} \int_A \int_A \rho^* {\rho^*}^\prime
\frac{(x-x^\prime)^i(x-x^\prime)^j(x-x^\prime)^k(x-x^\prime)^l}
{|{\bf x} - {\bf x}^\prime |^5} d^3x d^3x^\prime \,,
\nonumber \\
{\cal H}_A^{ij} &\equiv& \int_A \int_A \rho^* {\rho^*}^\prime 
\frac{{v^\prime}^i (x-x^\prime)^j}{|{\bf x} - {\bf x}^\prime |^3} d^3x
d^3x^\prime \,,
\nonumber \\
{\cal K}_A^{ij}&\equiv& \int_A \int_A \rho^* {\rho^*}^\prime
\frac{{\bf v}^\prime \cdot ({\bf x} - {\bf
x}^\prime)(x-x^\prime)^i(x-x^\prime)^j}{|{\bf x} - {\bf x}^\prime |^5} d^3x
d^3x^\prime \,,
\nonumber \\
{\cal K}_A^{ijkl}&\equiv& \int_A \int_A \rho^* {\rho^*}^\prime
\frac{{v^\prime}^i (x-x^\prime)^j(x-x^\prime)^k(x-x^\prime)^l}
{|{\bf x} - {\bf x}^\prime |^5} d^3x d^3x^\prime \,,
\nonumber \\
{{\cal K}_A^*}^{ijkl}&\equiv& \int_A \int_A \rho^* {\rho^*}^\prime
\frac{{\bf v}^\prime \cdot ({\bf x} - {\bf
x}^\prime)(x-x^\prime)^i(x-x^\prime)^j(x-x^\prime)^k(x-x^\prime)^l}
{|{\bf x} - {\bf x}^\prime |^7} d^3x
d^3x^\prime \,.
\eea

\section{Extracting total time derivatives}
\label{app:timederiv}

Using the Newtonian equations of motion plus the 1PN spin-orbit terms, it is
straightforward to establish a number of  identities, which may be used to
extract time derivatives from 2.5PN and 3.5PN terms.
For any non-negative integers $s$, $p$ and $q$, we obtain
\bea
\frac{d}{dt} \left ( \frac{v^{2s} {\dot r}^p}{r^q} \right ) &=&
\frac{v^{2s-2} {\dot r}^{p-1}}{r^{q+1}}
\left \{ pv^4 - (p+q)v^2{\dot r}^2 - 2s{\dot r}^2\frac{m}{r}
-pv^2\frac{m}{r} + \frac{p}{2}\frac{v^2}{r^3} {\bf {\tilde L}}_{\rm N} \cdot
(4 {\bf {\cal S}} + 3 \fourvec{\xi}) \right \}
\,,
\nonumber
\\
\frac{d}{dt} \left ( \frac{v^{2s} {\dot r}^p}{r^q} {\bf {\tilde L}}_{\rm N}
\right ) &=&
\frac{v^{2s-2} {\dot r}^{p-1}}{r^{q+1}}
\left \{
\left [  pv^4 - (p+q)v^2{\dot r}^2 - 2s{\dot r}^2\frac{m}{r}
-pv^2\frac{m}{r} + \frac{p}{2}\frac{v^2}{r^3} {\bf {\tilde L}}_{\rm N} \cdot
(4 {\bf {\cal S}} + 3 \fourvec{\xi}) \right ] {\bf {\tilde L}}_{\rm N}
\right .
\nonumber \\
&&
\left .
-\frac{v^2{\dot r}}{r} {\bf n} \times \left ( \left [ {\bf v} - \frac{3}{2}
{\dot r} {\bf n} \right ] \times (4 {\bf {\cal S}} + 3 \fourvec{\xi}) 
\right )
\right \} \,.
\label{timederiv1}
\eea
Another set of identities, to be used only in 3.5PN terms, require only the 
Newtonian equations of motion:
\bea
\frac{d}{dt} \left ( \frac{v^{2s} {\dot r}^p}{r^q} x^ix^j \right ) &=&
\frac{v^{2s-2} {\dot r}^{p-1}}{r^{q+1}}
\left \{  \left [pv^4 - (p+q)v^2{\dot r}^2 - 2s{\dot r}^2\frac{m}{r}
-pv^2\frac{m}{r} \right ] x^ix^j
\right .
\nonumber \\
&&
\left .
+2 v^2{\dot r} r x^{(i}v^{j)} \right \} \,,
\nonumber \\
\frac{d}{dt} \left ( \frac{v^{2s} {\dot r}^p}{r^q} v^iv^j \right ) &=&
\frac{v^{2s-2} {\dot r}^{p-1}}{r^{q+1}}
\left \{  \left [pv^4 - (p+q)v^2{\dot r}^2 - 2s{\dot r}^2\frac{m}{r}
-pv^2\frac{m}{r} \right ] v^iv^j
\right .
\nonumber \\
&&
\left .
-2m \frac{v^2{\dot r}}{r^2} x^{(i}v^{j)} \right \} \,,
\nonumber \\
\frac{d}{dt} \left ( \frac{v^{2s} {\dot r}^p}{r^q} x^iv^j \right ) &=&
\frac{v^{2s-2} {\dot r}^{p-1}}{r^{q+1}}
\left \{  \left [pv^4 - (p+q)v^2{\dot r}^2 - 2s{\dot r}^2\frac{m}{r}
-pv^2\frac{m}{r} \right ] x^iv^j
\right .
\nonumber \\
&&
\left .
+ v^2{\dot r} r \left ( v^iv^j - \frac{m}{r} n^in^j \right ) \right \} \,.
\label{timederiv2}
\eea

\section{Total energy and angular momentum}
\label{app:EJ}

In this appendix, we express the 
global definitions of mass, momentum, angular momentum
and center of mass in a PN expansion, including spin contributions.  
In Paper I, we defined the ``source moments''
\begin{subequations}
\begin{eqnarray}
P^\mu &\equiv&  \int_{\cal M} \tau^{\mu 0} d^3x \,, \\
{\cal I}^Q &\equiv& \int_{\cal M} \tau^{00} x^Q d^3x \,,
\\
{\cal J}^{iQ} &\equiv& \epsilon^{iab}\int_{\cal M}
x^{aQ} \tau^{0b} d^3x \,,
\\
{\cal M}^{ijQ}  &\equiv& \int_{\cal M} \tau^{ij}
{x}^{Q} d^3 x \,,
\end{eqnarray}
\label{sourcemoments}
\end{subequations}
where the capitalized superscript $Q$ denotes a multi-index 
($x^Q = x^{i_1} x^{i_2} \dots  x^{i_q}$). 
The integrals are to be taken over a
constant-time hypersurface $\cal M$, which lies within the near zone, i.e.
within one gravitational wavelength of the sources.
The quantities $P^0$ and $P^i$ are
naturally interpreted as the total mass-energy and momentum of the system,
respectively, $I^i \equiv {\cal I}^i \,(q=1)$ is the system dipole moment, 
which relates the center of
mass to the origin of coordinates, and $J^i \equiv {\cal J}^i \, (q=0)$ is the 
total angular momentum.  These quantities are conserved up to changes
induced by a flux of gravitational radiation to infinity, that is,
\begin{eqnarray}
{\dot P}^\mu &=& - \oint_{\partial {\cal M}} \tau^{\mu j}
d^2 S_j \,, \nonumber \\
{\dot J}^i &=&  - \epsilon^{iab} \oint_{\partial {\cal M}}
\tau^{jb} x^a d^2 S_j \,, \nonumber \\
{\dot I}^i &=& P^i - \oint_{\partial {\cal M}}
\tau^{0j} x^i d^2 S_j \,.
\end{eqnarray}
where $\partial {\cal M}$ denotes the boundary of $\cal M$.  
Thus, these quantities are conserved and well-defined only up to and
including 2PN order.

We now wish to evaluate these conserved quantities explicitly to 1PN order
for a many-body system.  Explicit expressions for $\tau^{\alpha\beta}$
through the relevant order may be obtained from Eqs. (5.7) and (5.9a) of
Paper I, together with Eqs. (2.21) of Paper II which serve to
convert expressions to dependence on $\rho^*$.  They are given by
\bes
\bea
\tau^{00} &=&  \rho^* \left ( 1 + \frac{1}{2}v^2 - \frac{1}{2} U 
+ \frac{3}{8} v^4 + v^2 U + \frac{3}{2} U^2 
+ \frac{1}{2} \Phi_1 - \Phi_2 - 2 v^j V^j + \frac{1}{4} v^j {\dot X}^{,j} 
\right ) \,,
\label{tau00}
\\
\tau^{0j} &=& \rho^* v^j \left ( 1 + \frac{1}{2}v^2 - U \right )
-\frac{1}{2} {\dot X}^{,j} 
+ \frac{1}{4\pi} \nabla^k \left ( 8UV^{[k,j]} + 4U{\dot U}\delta^{jk}
- {\dot X}^{,(j} U^{,k)} + \frac{1}{2}\delta^{jk} {\dot X}^{,m} U^{,m}
\right )
\,,
\\
\tau^{ij} &=& \rho^* v^i v^j  + \frac{1}{4\pi} \left ( U^{,i} U^{,j} - 
\frac{1}{2} \delta^{ij} \nabla U^2 \right ) \,,
\eea
\label{taumunu}
\ees
where we have extracted total divergences of functions, such as $\nabla \cdot
(U \nabla U)$ from the expression for $\tau^{00}$, 
since they will not affect the integral quantities $P^0$ and $I^i$, to the
orders considered.

Expanding potentials about the baryonic center of mass of each body as
described in Sec. (\ref{sec2.5pnA}), 
keeping spin terms and other
``internal'' terms, but discarding terms that vanish as $R^2$ and
higher as the body's size tends to zero, and applying the virial relations
of Appendix \ref{app:virial}, we obtain the same expressions (\ref{EJPN}) 
for $E$, ${\bf J}$, and ${\bf I}$, as
we found from the equations of motion.

\end{document}